\documentclass[prb,twocolumn,showkeys,showpacs,preprintnumbers,amsmath,amssymb,floatfix,superscriptaddress]{revtex4}

\usepackage{graphicx}
\usepackage{dcolumn}
\usepackage{bm}
\usepackage{dsfont}
\usepackage{amsmath}
\usepackage{amssymb}
\usepackage{soul}
\renewcommand{\vec}[1]{\boldsymbol{#1}}

\begin{document}

\title{Angle-Dependent Spin-Wave Resonance Spectroscopy of (Ga,Mn)As Films}

\author{L. Dreher}
\email{dreher@wsi.tum.de} 
\affiliation{Walter Schottky Institut, Technische Universit\"at M\"unchen, Am Coulombwall 4, 85748 Garching, Germany}
\author{C. Bihler}
\affiliation{Walter Schottky Institut, Technische Universit\"at M\"unchen, Am Coulombwall 4, 85748 Garching, Germany}
\author{E. Peiner}
\affiliation{Institut f\"ur Halbleitertechnik, Technische Universit\"at Braunschweig,Hans-Sommer-Stra\ss e 66, 38023 Braunschweig, Germany}
\author{A. Waag}
\affiliation{Institut f\"ur Halbleitertechnik, Technische Universit\"at Braunschweig,Hans-Sommer-Stra\ss e 66, 38023 Braunschweig, Germany}
\author{W. Schoch}
\affiliation{Institut f\"ur Quantenmaterie, Universit\"at Ulm, 89069 Ulm, Germany}
\author{W. Limmer}
\affiliation{Institut f\"ur Quantenmaterie, Universit\"at Ulm, 89069 Ulm, Germany}
\author{S.T.B. Goennenwein}
\affiliation{Walther-Mei\ss ner-Institut, Bayerische Akademie der Wissenschaften, Walther-Mei\ss ner-Stra\ss e 8, 85748 Garching, Germany}
\author{M.S. Brandt}
\affiliation{Walter Schottky Institut, Technische Universit\"at M\"unchen, Am Coulombwall 4, 85748 Garching, Germany}

\date{\today}

\begin{abstract}
A modeling approach for standing spin-wave resonances based on a finite-difference formulation of the Landau-Lifshitz-Gilbert equation is presented. In contrast to a previous study [Bihler {\it et al.}, Phys.~Rev.~B {\bf 79}, 045205 (2009)], this formalism accounts for elliptical magnetization precession and magnetic properties arbitrarily varying across the layer thickness, including the magnetic anisotropy parameters, the exchange stiffness, the Gilbert damping, and the saturation magnetization. To demonstrate the usefulness of our modeling approach, we experimentally study a set of (Ga,Mn)As samples grown by low-temperature molecular-beam epitaxy by means of angle-dependent standing spin-wave resonance spectroscopy and electrochemical capacitance-voltage measurements. By applying our modeling approach, the angle dependence of the spin-wave resonance data can be reproduced in a simulation with one set of simulation parameters for all external field orientations. We find that the approximately linear gradient in the out-of-plane magnetic anisotropy is related to a linear gradient in the hole concentrations of the samples.
\end{abstract}


\maketitle

\section{Introduction}\label{introduction}

Due to their particular magnetic properties, including magnetic anisotropy,\cite{PRB80_155203,APL86_112512,PRB81_245202} anisotropic magneto-resistance \cite{PRL99_147207,PRB77_205210} and magneto-thermopower,\cite{PRL97_36601} in the past years ferromagnetic semiconductors have continued to be of great scientific interest in exploring new physics and conceptual spintronic devices.\cite{RMP78_809,NM9_952,NM9_965,NM9_898,PRL106_57204} The most prominent ferromagnetic semiconductor is (Ga,Mn)As, where a small percentage of Mn atoms on Ga sites introduces localized magnetic moments as well as itinerant holes which mediate the ferromagnetic interaction of the Mn spins ($p$-$d$ exchange interaction).\cite{PRB63_195205}
Both theoretical and experimental studies have shown that the magnetic anisotropy, i.e., the dependence of the free energy of the ferromagnet on the magnetization orientation, depends on the elastic strain and the hole concentration in the (Ga,Mn)As layer,\cite{PRB63_195205,PRB79_195206} opening up several pathways to manipulate the magnetic anisotropy of (Ga,Mn)As.\cite{PSS2_96,NJoP10_65003,PRB78_45203}

A common spectroscopic method to probe the magnetic anisotropy of ferromagnets and in particular (Ga,Mn)As, is angle-dependent ferromagnetic resonance (FMR),\cite{JAP32_129,JoMaMM250_164,JAP91_7484,PRB67_205204,APL89_012507,LiuFurdyna_FMR,PRB77_165204} where FMR spectra are taken as a function of the orientation of the external magnetic field. If the magnetic properties of the ferromagnet are homogeneous, a zero wave vector ($k=0$) mode of collectively, uniformally precessing magnetic moments couples to the microwave magnetic field, e.g., in a microwave cavity, allowing for a detection of the magnetization precession. The resonance field of this mode, referred to as uniform resonance magnetic field, depends on the employed microwave frequency and the magnetic anisotropy parameters. Thus, by recording FMR spectra at different orientations of the external field with respect to the crystal axes, the anisotropy parameters can be deduced from the experiment. However, if the magnetic properties of a ferromagnetic layer are non-homogeneous or the spins at the surface and interface of the layer are pinned, non-propagating modes with $k\neq 0$, referred to as standing spin-wave resonances (SWR), can be excited by the cavity field and thus be detected in an FMR experiment. On one hand this can hamper the derivation of anisotropy parameters, on the other hand a detailed analysis of these modes can elucidate the anisotropy profile of the layer and the nature of spin pinning conditions. Furthermore, the excitation of spin waves is of topical interest in combination with spin-pumping,\cite{APL97_252504,PRL106_216601,P4_40,PRL107_46601} i.e., the generation of pure spin currents by a precessing magnetization.\cite{PRL88_117601,PRL104_46601,RMP77_1375} In this context, the exact knowledge of the magnetization precession amplitude as a function of the position coordinate within the ferromagnet is of particular importance.\cite{APL97_252504}

Several publications report on SWR modes in (Ga,Mn)As with a mode spacing deviating from what is expected according to the Kittel model for magnetically homogeneous films with pinned spins at the surface.\cite{APL82_730,JoS16_143,PRB69_125213,MITo43_3019,PRB75_195220,PRB79_45205} These results have been attributed to an out-of-plane anisotropy field linearly\cite{APL82_730,PRB79_45205} or quadratically varying\cite{PRB69_125213,MITo43_3019,PRB75_195220} as a function of the depth into the layer, as well as to specific spin pinning conditions at the surface and at the interface to the substrate.\cite{PRB75_195220} While most of these studies have focused on the spacings of the resonance fields when modeling SWR measurements, in Ref.~\onlinecite{PRB79_45205} a more sophisticated approach, based on a normal mode analysis,\cite{JAP48_382,Hoekstra_diss} was employed to model resonance fields as well as relative mode intensities for the external field oriented along high-symmetry directions, assuming a circularly precessing magnetization.

In this work, we present a more general modeling approach for SWR, based on a finite-difference formulation of the Landau-Lifshitz-Gilbert (LLG) equation. This approach holds for any orientation of the external magnetic field and accounts for elliptical magnetization precession [Sec.~\ref{theory}]. It allows for a simulation of arbitrarily varying profiles of the magnetic properties across the thickness of the film, including vatiations of the magnetic anisotropy parameters, the exchange stiffness, and the Gilbert damping parameter. As the result of the simulation, we obtain the Polder susceptibility tensor as a function of the depth within the ferromagnet. Based on this result, the absorbed power upon spin wave resonance and the magnetization precession amplitude as a function of the depth can be calculated for any orientation of the external magnetic field.

We apply our modeling approach to a set of four (Ga,Mn)As samples epitaxially grown with different V/III flux ratios [Sec.~\ref{experiment}], motivated by the observation that V/III flux ratios of $\lesssim 3$ lead to a gradient in the hole concentration $p$ [Ref.~\onlinecite{PRB71_205213}], which in turn is expected to cause non-homogeneous magnetic anisotropy parameters.\cite{APL82_730,PRB79_45205} Electrochemical capacitance-voltage (ECV) measurements revealed a nearly linear gradient in $p$ across the thickness of the layers investigated. To show that our modeling approach is capable of simulating SWR spectra for arbitrary magnetic field orientations, angle-dependent SWR data were taken and compared with the model using one set of magnetic parameters for each sample, revealing gradients in the uniform resonance magnetic fields. We discuss the influence of the gradient in $p$ on the observed uniform resonance field gradients as well as possible influences of strain and saturation magnetization gradients on the observed out-of-plane anisotropy profile. It should be emphasized, however, that the objective of this work is to show the usefulness of our modeling approach, while a detailed investigation of the origin of the gradient in the out-of-plane magnetic anisotropy profile and therefore a detailed understanding of the particular materials physics of (Ga,Mn)As is beyond the scope of this study.
Finally, we summarize our results and discuss further potential applications of this work [Sec.~\ref{summary}].

\section{Theoretical Considerations}\label{theory}
\begin{figure}
\includegraphics[]{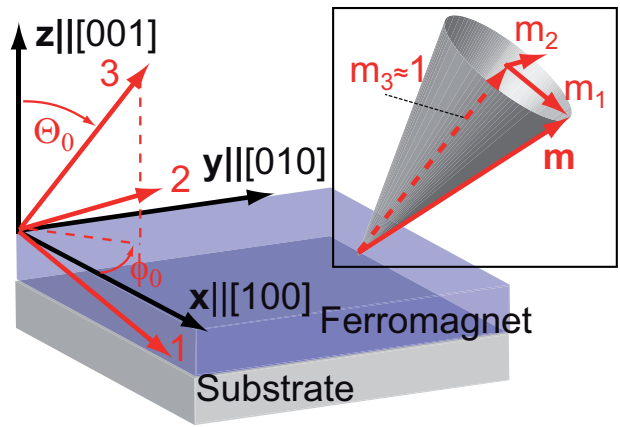}
\caption{(color online) Relation between the two coordinate systems employed. The $(x,y,z)$ frame of reference is spanned by the cubic crystal axes, while the $(1,2,3)$ coordinate system is determined by the equilibrium orientation of the magnetization (3-direction) and two transverse directions, the 2-direction being parallel to the film plane; the latter system is $z$ and $\mu_0\boldsymbol{H}$ dependent, as described in the text.}
\label{fig:SWR_CoordinateSystem}
\end{figure}

In this section, we provide the theoretical framework necessary to describe the full angle dependence of the spin-wave resonance spectra presented in Sec.~\ref{experiment}.
Referring to the coordinate system depicted in Fig.~\ref{fig:SWR_CoordinateSystem}, we start from the canonical expression for the free enthalpy density (normalized to the saturation magnetization $M$) for a tetragonally distorted (Ga,Mn)As film\cite{RoPiP61_755,PRB67_205204,PRB74_205205,PRB79_195206}
\begin{eqnarray}
    G&=& \mathrm{const} -\mu_0\boldsymbol{H}\cdot\boldsymbol{m}+ B_{001}m_z^2+B_{4\perp}m_z^4\nonumber\\ 
    					 &+& B_{4\parallel}(m_x^4+m_y^4)+{\frac{1}{2}}B_{1\bar{1}0}(m_x-m_y)^2\label{eq:FE_001}.
\end{eqnarray}
Here, $\mu_0\boldsymbol{H}$ is a static external magnetic field, $B_{001}$ is a uniaxial out-of-plane anisotropy parameter, reflecting shape and second-order crystalline anisotropy,\cite{PRB79_195206} $B_{4\perp}$, $B_{4\parallel}$, and $B_{1\bar{1}0}$ are fourth-order crystalline and second-order uniaxial in-plane anisotropy parameters, respectively;\cite{PRB80_155203} $m_x$, $m_y$, $m_z$ denote the components of the normalized magnetization vector $\boldsymbol{m}(z)=\boldsymbol{M}(z)/M(z)$ along the cubic axes [100], [010], and [001], respectively. We assume the magnetic properties of the layer to be homogeneous laterally (within the $xy$ plane) and inhomogeneous vertically (along $z$); the anisotropy parameters in Eq.~\eqref{eq:FE_001} and the magnetization are consequently a function of the spatial variable $z$. To obtain the anisotropy parameters from Eq.~\eqref{eq:FE_001} in units of energy density, it would therefore be required to know the $z$ dependence and the absolute value of $M$.

The minimum of Eq.~\eqref{eq:FE_001} determines the equilibrium orientation of the magnetization, given by the angles $\theta_0=\theta_0(z)$ and $\phi_0=\phi_0(z)$, cf.~Fig.~\ref{fig:SWR_CoordinateSystem}. To describe the magnetization dynamics, we introduce a new frame of reference $(1,2,3)$ shown in Fig.~\ref{fig:SWR_CoordinateSystem}, in which the equilibrium orientation of the magnetization $\vec{m_0}$ coincides with the axis 3. For small perturbations, the magnetization precesses around its equilibrium with finite transverse components of the magnetization $m_i$ ($i=1,2$) as illustrated in the inset in Fig.~\ref{fig:SWR_CoordinateSystem}. The transformation between the two coordinate systems is given in the Appendix \ref{Appendix_A_SWR} by Eqs.~\eqref{eq:SWR_Transformation} and \eqref{eq:SWR_TransformationMatrix}. We write for the (normalized) magnetization

\begin{equation}
\vec{m}=\underbrace{\left(\begin{array}{c}
	0\\
	0\\
	1\\
\end{array}\right)}_{\vec{m_0}}
+\left(\begin{array}{c}
	m_1\\
	m_2\\
0\\
\end{array}\right)+O(m_1^2,m_2^2).
\label{eq:SWR_magnetization}
\end{equation}

The evolution of the magnetization under the influence of an effective magnetic field $\mu_0 \boldsymbol{H}_{\rm{eff}}$ is described by the LLG equation\cite{LandauLifshitz,gilbert_phenomenological_2004}
\begin{equation}
\partial_t\boldsymbol{m}=-\gamma \boldsymbol{m}\times\mu_0 \boldsymbol{H}_{\rm{eff}}+\alpha \boldsymbol{m}\times \partial_t \boldsymbol{m},
\label{eq:SWR_LLG}
\end{equation}
where $\gamma$ is the gyromagnetic ratio and $\alpha$ a phenomenological damping parameter. The effective magnetic field is given by\cite{PRB79_45205}
\begin{equation}
\mu_0 \boldsymbol{H}_{\rm{eff}}=-\nabla_{\boldsymbol{m}}G +\frac{D_\text{s}}{M} \nabla^2 \boldsymbol{M}+\mu_0 \boldsymbol{h}(t),
\label{eq:H_eff}
\end{equation}
where $\nabla_{\boldsymbol{m}}=(\partial_{m1},\partial_{m2},\partial_{m3})$ is the vector differential operator
with respect to the components of $\boldsymbol{m}$, $D_\text{s}=2A/M$ is the exchange stiffness with the exchange constant $A$, $\nabla^2$ is the spatial differential operator $\nabla^2 =\partial_x^2+\partial_y^2+\partial_z^2$, and $\boldsymbol{h}(t)=\boldsymbol{h}_0e^{-i\omega t}$ is the externally applied microwave magnetic field with the angular frequency $\omega$; $\boldsymbol{h}(t)$ is oriented perpendicularly to $\mu_0\boldsymbol{H}$. Since the magnetic properties are independent of $x$ and $y$, Eq.~\eqref{eq:SWR_LLG} simplifies to
\begin{equation}
\partial_t\boldsymbol{m}=-\gamma \boldsymbol{m}\times \left[-\nabla_{\boldsymbol{m}}G +D_\text{s} \boldsymbol{m}''+ \mu_0\boldsymbol{h}(t) \right]+\alpha \boldsymbol{m}\times \partial_t \boldsymbol{m},
\label{eq:SWR_LLG_with_Heff}
\end{equation}
with $\boldsymbol{m}''=\partial_z^2 \boldsymbol{m}$, neglecting terms of the order of $m_i^2$ (for $i=1,2$).
By definition of the $(1,2,3)$ coordinate system, the only non-vanishing component of $\nabla_{\boldsymbol{m}}G$ in the equilibrium is along the $3$-direction. For small deviations of $\vec{m}$ from the equilibrium we find\cite{PRB38_2237}
\begin{equation}
\nabla_{\boldsymbol{m}}G=\left(\begin{array}{c}
     G_{11}m_1+G_{21}m_2 \\
     G_{12}m_1+G_{22}m_2\\
     G_3\\
     \end{array}\right),
\label{eq:SWR_Expansion_G}
\end{equation}
where we have introduced the abbreviations $G_i=\partial_{m_i}G|_{\vec{m}=\vec{m_0}}$ and $G_{ij}=\partial_{m_i}\partial_{m_j}G|_{\vec{m}=\vec{m_0}}$; the explicit expressions for these derivatives are given in the Appendix \ref{Appendix_A_SWR}.

In the following, we calculate the transverse magnetization components assuming a harmonic time dependence $m_i=m_{i,0}e^{-i\omega t}$. The linearized LLG equation, considering only the transverse components, reads as 
\begin{equation}
    \left(\begin{array}{cc}
    H_{11} & H_{12} \\
     H_{21} & H_{22}\\
     \end{array}\right)
     \left(\begin{array}{c}
     m_1 \\
     m_2\\
     \end{array}\right)
     - D_\text{s}\left(\begin{array}{c}
     m_1'' \\
     m_2''\\
     \end{array}\right)
     =
     \mu_0 
     \left(\begin{array}{c}
     h_{1} \\
     h_{2}\\
     \end{array}\right),
\label{eq:SpinWaveEquationElliptical}
\end{equation}
where we have introduced the abbreviations $H_{11}=G_{11}-G_3-i\alpha\omega/\gamma$, $H_{12}=H_{21}^*=G_{12}+i\omega/\gamma$, and $H_{22}=G_{22}-G_3-i\alpha\omega/\gamma$.
We have dropped all terms which are non-linear in $m_i$ and products of $m_i$ with the driving field.

Resonant uniform precession of the magnetization ($m_i''=0$) occurs at the so called uniform resonance field $\mu_0H_{\mathrm{uni}}(z)$, which is found by solving the homogeneous ($\boldsymbol{h}=0$) equation
\begin{eqnarray}
   H_{11}(z)H_{22}(z)-H_{12}(z)H_{21}(z)&=&0\nonumber\\
 \Leftrightarrow  (G_{11}-G_3)(G_{22}-G_3)-G_{12}^2&=&\left(\frac{\omega}{\gamma}\right)^2
\label{eq:UniformResonanceField}
\end{eqnarray}
for $\mu_0 H$, neglecting the Gilbert damping ($\alpha=0$). Equation \eqref{eq:UniformResonanceField} can be used to derive anisotropy parameters from angle-dependent FMR spectra. As extensively discussed by Baselgia et al.\cite{PRB38_2237}, using Eq.~\eqref{eq:UniformResonanceField} is equivalent to using the method of Smit and Beljers, which employs second derivatives of the free enthalpy with respect to the spherical coordinates.\cite{PRR10_113,PRB74_205205,PRB81_205210}

To illustrate the role of the uniform resonance field in the context of spin-wave resonances, we consider the special case where magnetization is aligned along the [001] crystal axis ($\theta_0=0$), before we deal with the general case of arbitrary field orientations.
Neglecting the uniaxial in-plane anisotropy ($B_{1\bar{1}0}=0$) since this anisotropy is typically weaker than all other anisotropies,\cite{PRB74_205205,PRB79_195206} we find $G_3=-\mu_0H+2B_{001}+4B_{4\perp}$ and $G_{11}=G_{22}=G_{12}=0$, resulting in the uniform resonance field
\begin{equation}
\mu_0 H_\mathrm{uni}^{001}(z)=\omega/\gamma+2B_{001}(z)+4B_{4\perp}(z).
\label{eq:SWR_uniformResonace001}
\end{equation}
To find the eigenmodes of the system, we consider the unperturbed and undamped case, i.e., $\alpha=0$ and $\boldsymbol{h}=0$ in Eq.~\eqref{eq:SpinWaveEquationElliptical}. With $m_2=i m_1=\tilde{m}$ we find the spin-wave equation 
\begin{equation}
D_\text{s} \tilde{m}''+\mu_0 H_\mathrm{uni}^{001}(z) \tilde{m}=\mu_0H \tilde{m}
\label{eq:SpinWaveEquationSimple}
\end{equation}
in agreement with Ref.~\onlinecite{PRB79_45205}.
\begin{figure}[h]
\includegraphics[]{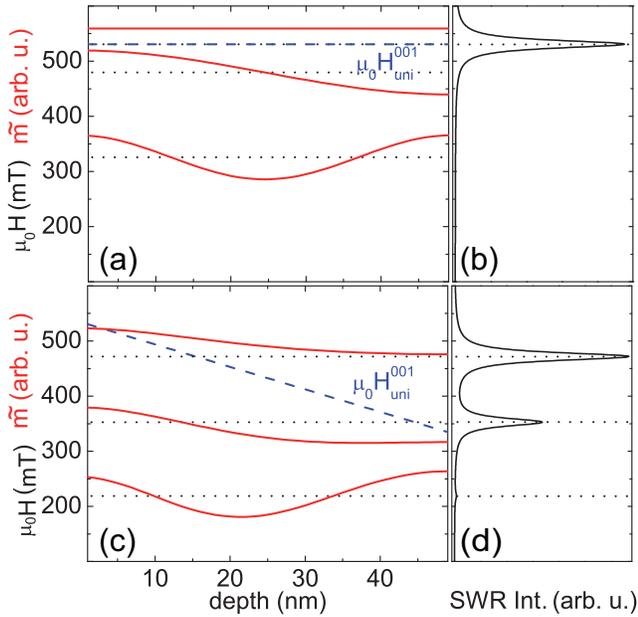}
\centering
\caption{Simulation to demonstrate the influence of the uniform resonance field $\mu_0 H_\mathrm{uni}^{001}$ on the SWR modes for $\boldsymbol{m}_0||[001]$, assuming circular precession. In (a), $\mu_0 H_\mathrm{uni}^{001}$ is set to be constant across the layer, while in (c) it varies linearly (blue, dashed lines), in analogy to a square potential and a triangular potential, respectively. The dotted black lines are the resonance fields, calculated assuming boundary conditions of natural freedom, see text. The solid red lines show the eigenmodes of the system, i.e., the precession amplitude $\tilde{m}$ of the magnetization; for each mode the dotted line corresponds to $\tilde{m}=0$. As can be seen in (a), for a constant uniform resonance field the first mode occurs at the uniform resonance field and exhibits a constant precession amplitude across the layer, i.e., an FMR mode. The second and third mode (higher-order modes are not shown) exhibit a non-uniform magnetization profile. In order to couple to the driving field the modes need to have a finite net magnetic moment. As can be seen in (a), the positive and negative areas of the second and third mode are equal, thus these modes are not visible in the SWR spectrum (b). This is in contrast to the case of the linearly varying uniform resonance field (c) where the mode profile is given by Airy functions, which have a nonzero net magnetic moment also for the second and third mode, resulting in a finite SWR intensity of these modes (d). The spectra in (b) and (d) were calculated by integrating over the eigenmodes $\tilde{m}$ and convoluting the square of the result with Lorentzians.}
\label{fig:DidaktikSWR}
\end{figure}%
The relation of the anisotropy parameters defined in Ref.~\onlinecite{PRB79_45205} to the ones used here is given by $B_{001}\!=\!K_\mathrm{eff}^{100}/M+B_{1\bar{1}0}$, $B_{1\bar{1}0}\!=\!-K_\mathrm{u}^{011}/M$, $2B_{4\perp}\!=\!-K_\mathrm{c1}^{\perp}/M$, and $2B_{4\parallel}\!=\!-K_\mathrm{c1}^{\parallel}/M$. Equation \eqref{eq:SpinWaveEquationSimple} is mathematically equivalent to the one-dimensional time-independent Schr\"odinger equation, where the uniform resonance field corresponds to the potential, $\tilde{m}$ to the wave function, $\mu_0H$ to the energy, and $D_\text{s}$ is proportional to the inverse mass. To calculate the actual precession amplitude of the magnetization, the coupling of the eigenmodes of Eq.~\eqref{eq:SpinWaveEquationSimple} to the driving field is relevant, which is proportional to the net magnetic moment of the mode.\cite{Hoekstra_diss,PRB79_45205} In analogy to a particle in a box, the geometry of the uniform resonance field as well as the boundary conditions determine the resonance fields and the spatial form of the precession amplitude. For the remainder of this work, we assume the spins to exhibit {\it natural freedom} at the boundaries of the film, i.e., $\partial_z \tilde{m}= \tilde{m}'=0$ at the interfaces,\cite{PiSS9_191,PRB79_45205} since these boundary conditions have been shown to describe the out-of-plane SWR data of similar samples well.\cite{PRB79_45205} To graphically illustrate the influence of the uniform resonance field on the SWR modes, we consider in Fig.~\ref{fig:DidaktikSWR} a ferromagnetic layer with a thickness of 50~nm with constant magnetic properties across the layer (a) and with a linearly varying uniform resonance field (c); in both cases we assume $D_\text{s}=13~\text{Tnm}^{-2}$, a similar value as obtained in previous studies.\cite{PRB79_45205} For these conditions, we numerically solve Eq.~\eqref{eq:SpinWaveEquationSimple} by the finite difference method described in the Appendix \ref{subsec:NumericallySWR_simple}, in order to obtain the resonance fields (eigenvalues) and the $z$ dependence of the transverse magnetic moments (eigenfunctions). To which amount a mode couples to the driving field is determined by the net magnetic moment of the mode, which is found by integrating $\tilde{m}(z)$ over the thickness of the film. For the magnetically homogeneous layer, the only mode that couples to the driving field is the uniform precession mode at $\mu_0 H_\mathrm{uni}^{001}$, since modes of higher order have a zero net magnetic moment [Fig.~\ref{fig:DidaktikSWR} (a)], resulting in one resonance at the uniform resonance field, cf.~Fig.~\ref{fig:DidaktikSWR} (b). For the non-uniform layer, with $\mu_0 H_\mathrm{uni}^{001}(z)$ linearly varying across the film, the mode profile is given by Airy functions\cite{Hoekstra_diss,APL82_730,PRB79_45205} and various non-uniform modes couple to the driving field, resulting in several spin-wave resonances with their amplitude proportional to the square of the net magnetic moment\cite{PRB79_45205,Hoekstra_diss} of the corresponding mode, cf.~Fig.~\ref{fig:DidaktikSWR} (c) and (d).\\

We now turn to the general case of arbitrary field orientations. Due to the magnetic anisotropy profile, the magnetization orientation is a priori unknown and a function of $z$ and $\mu_0\vec{H}$. Furthermore, the assumption of a circularly precessing magnetization is not generally justified.
To solve Eq.~\eqref{eq:SpinWaveEquationElliptical} for arbitrary field orientations, we employ a finite difference method as outlined in the Appendix \ref{subsec:NumericallySWR}. By solving Eq.~\eqref{eq:SpinWaveEquationElliptical}, we obtain the $z$ dependent generalized Polder susceptibility tensor $\bar{\chi}(\mu_0 \boldsymbol{H},z)$, which relates the transverse magnetization components $M_i(z)=M(z)m_i(z)$ with the components of the driving field by
\begin{equation}
 \left(\begin{array}{c}
   M_1 \\
   M_2\\
   \end{array}\right)=\bar{\chi}(\mu_0 \boldsymbol{H},z) \left(\begin{array}{c}
   h_{1} \\
   h_{2}\\
   \end{array}\right).
\label{eq:Polder}
\end{equation}
In a microwave absorption measurement, the components $M_i$ which are out-of-phase with the driving field are detected. The absorbed power density is related to the imaginary part of $\bar{\chi}(\mu_0 \boldsymbol{H},z)$ and can be calculated by\cite{PRL106_117601}
\begin{equation}
 P=\frac{\omega\mu_0 }{2z_0}\mathrm{Im}\left\{ \int_{-z_0}^{0}\left[\left(\begin{array}{c}
   h_{1}^*, h_{2}^* \\
   \end{array}\right)\bar{\chi}(\mu_0 \boldsymbol{H},z) \left(\begin{array}{c}
   h_{1} \\
   h_{2}\\
   \end{array}\right)\right]\mathrm{d}z\right\},
\label{eq:absorbedPower}
\end{equation}
where $z_0$ is the thickness of the ferromagnetic layer. Note that the position coordinate $z$ is negative in the film, cf.~Fig.~\ref{fig:SWR_CoordinateSystem}.

\begin{figure*}
\includegraphics[width=\textwidth]{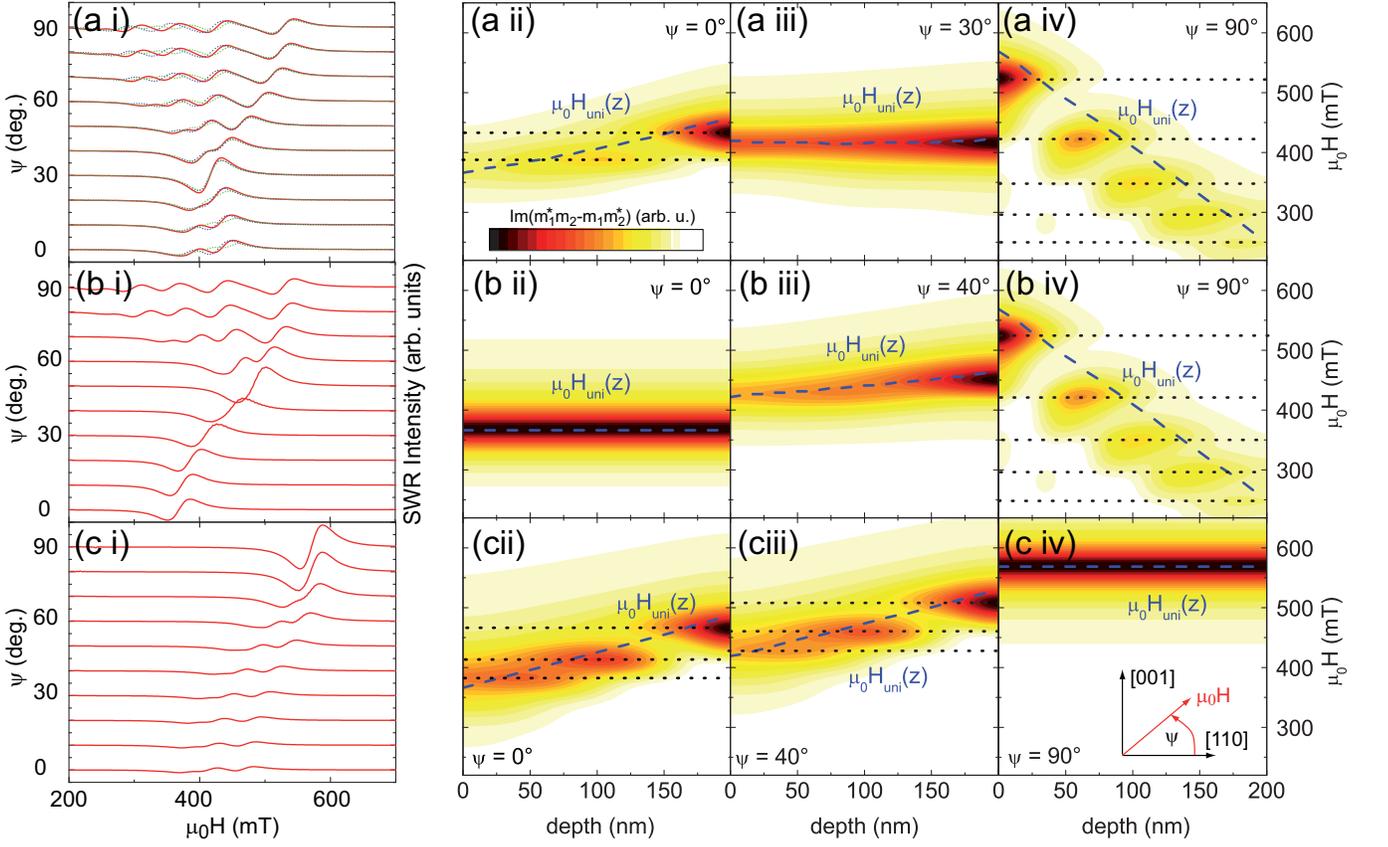}
\centering
\caption{Atlas illustrating the influence of gradients in the anisotropy parameters on SWR spectra. In (a) all anisotropy parameters are kept constant with the values given in the text, except $B_{001}$ which is varied linearly. Correspondingly, in (b) and (c) $B_{4\perp}$ and $B_{4||}$ were varied linearly, respectively. Panels (i) show the first derivative of simulations using Eq.~\eqref{eq:absorbedPower} with respect to $\mu_0 H$ and panels (ii)-(iv) show the precession cone $\text{Im}(m_1^*m_2-m_1m_2^*)$ in a color plot together with the uniform resonance field $\mu_0H_\text{uni}(z)$ (dashed blue lines) at three different external field orientations; the black dotted lines indicate the resonance field positions of the modes. Panel (a i) additionally shows the influence of a linear gradient in the exchange stiffness parameter on the spin-wave spectra, see text for further details and discussion.}
\label{fig:Bildatlas}
\end{figure*}

To obtain an impression of how gradients in different anisotropy parameters influence the SWR spectra, we plot in Fig.~\ref{fig:Bildatlas} simulated SWR spectra together with the magnetization precession cone as a function of depth in the ferromagnetic layer. We assume a constant saturation magnetization (its value is not relevant for the outcome of the simulation), a constant exchange stiffness $D_\text{s}=35$~Tnm$^2$ unless otherwise specified, $\alpha=0.09$, and $B_{001}=90$~mT, $B_{4||}=-50$~mT, $B_{4\perp}=15$~mT. In Fig.~\ref{fig:Bildatlas} (a), we assume $B_{001}$ to vary across the layer thickness according to $B_\mathrm{001}(z)= B_\mathrm{001}-b_\mathrm{001}\times z$ with $b_{001}=-0.8$~mT/nm. Figure \ref{fig:Bildatlas}~(a i) shows the simulated SWR spectra calculated by taking the first derivative of Eq.~\eqref{eq:absorbedPower} with respect to $\mu_0H$ for different angles $\psi$ defined in the inset in Fig.~\ref{fig:Bildatlas} (c iv). We observe several SWR modes for $\mu_0\vec{H}||[001]$ which become less as $\mu_0\vec{H}$ is tilted away from [001]. At $\psi=40^\circ$ only one mode is visible while for $\psi=0^\circ$ we again observe multiple SWR modes. This observation can be understood by considering the uniform resonance fields as a function of the depth for these orientations. In Fig.~\ref{fig:Bildatlas}~(a ii)-(a iv), we show the uniform resonance field (dashed blue line) for $\psi=0^\circ$, $\psi=30^\circ$, and $\psi=90^\circ$, respectively, together with the magnetization precession cone $\text{Im}(m_1^*m_2-m_1m_2^*)$ in a contour plot as a function of depth and $\mu_0H$. At $\psi=90^\circ$, the uniform resonance field varies strongly across the film, which can be understood by considering Eq.~\eqref{eq:SWR_uniformResonace001}. This results in several spin wave modes with their resonance fields indicated by dotted lines. 

For other field orientations, the formula for the uniform resonance field can also be derived but results in a longer, more complex equation than Eq.~\eqref{eq:SWR_uniformResonace001}. Important in this context is that positive values of $B_{001}$ lead to an increase (decrease) of the resonance field for the magnetization oriented perpendicular (parallel) to the film plane, accounting for the reversed sign of the slopes of $\mu_0H_\text{uni}$ in Fig.~\ref{fig:Bildatlas}~(a ii) and (a iv). Consequently, in between those two extreme cases $\mu_0H_\text{uni}$ must be constant across the layer for some field orientation, in our case for $\psi=30^\circ$, resulting in a single SWR mode, cf.~Fig.~\ref{fig:Bildatlas}~(a i) and (a iii). In addition to the SWR simulations with constant $D_\text{s}$, we plot in Fig.~\ref{fig:Bildatlas}~(a i) simulated SWR spectra with $D_\text{s}$ varying linearly across the film with $D_\text{s}=35-65$~Tnm$^2$ (blue, dotted lines) and $D_\text{s}=35-5$~Tnm$^2$ (green, dotted lines). A decreasing $D_\text{s}$ leads to a decreasing spacing in the modes and vice versa for an increasing $D_\text{s}$ as can be seen, e.g., for $\mu_0\vec{H}||[001]$. 

In Fig.~\ref{fig:Bildatlas}~(b), we consider the case where all magnetic parameters are constant with the values given above, except $B_\mathrm{4\perp}(z)=B_\mathrm{4\perp}-b_\mathrm{4\perp}\times z$ with $b_{4\perp}=-0.4$~mT/nm. As evident from Eq.~\eqref{eq:SWR_uniformResonace001}, this results in the same slope of $\mu_0H_\text{uni}$ for $\psi=90^\circ$ as in the case above where we varied $B_{001}$ only, cf.~Fig.~\ref{fig:Bildatlas}~(a iv) and (b iv). In contrast to the case depicted in (a), however, here for $\psi=0^\circ$ the uniform resonance field is constant. This can be understood when evaluating the parameters that enter in the calculation of the uniform resonance field [Eq.~\eqref{eq:UniformResonanceField}]. If $\vec{m}$ is in the film plane, none of the parameters in Eqs.~\eqref{eq:SWR_G_3}-\eqref{eq:SWR_G_22} depends on $B_{4\perp}$, resulting in a constant uniform resonance field for $\psi=0^\circ$. As $\vec{m}$ is tilted away from the film plane, $B_{4\perp}$ enters in some of the terms Eqs.~\eqref{eq:SWR_G_3}-\eqref{eq:SWR_G_22}. As a consequence, $\mu_0H_\text{uni}$ varies, first such that it increases [cf.~Fig.~\ref{fig:Bildatlas}~(b iii)] and finally, such that it decreases as a function of depth [cf.~Fig.~(b iv)].

Finally, we discuss the case where all parameters are constant except $B_\mathrm{4||}(z)=B_\mathrm{4||}-b_\mathrm{4||}\times z$ with $b_\mathrm{4||}=-0.4$~mT/nm [Fig.~\ref{fig:Bildatlas} (c)]. Here, $\mu_0H_\text{uni}$ is constant for $\psi=90^\circ$ as predicted by Eq.~\eqref{eq:SWR_uniformResonace001}. As $\vec{m}$ is tilted away from [001] a varying $B_{4||}$ leads to a varying uniform resonance field as shown in Fig. ~\ref{fig:Bildatlas} (c ii) and (c iii). Here, a sign reversal of the slope as it was the case in Fig.~\ref{fig:Bildatlas}~(a) and (b) does not take place and multiple resonances occur, starting from $\psi=60^\circ$ [Fig.~\ref{fig:Bildatlas} (c i)].

\section{Experimental Results and Discussion}\label{experiment}

\begin{figure}[h]
  \includegraphics[]{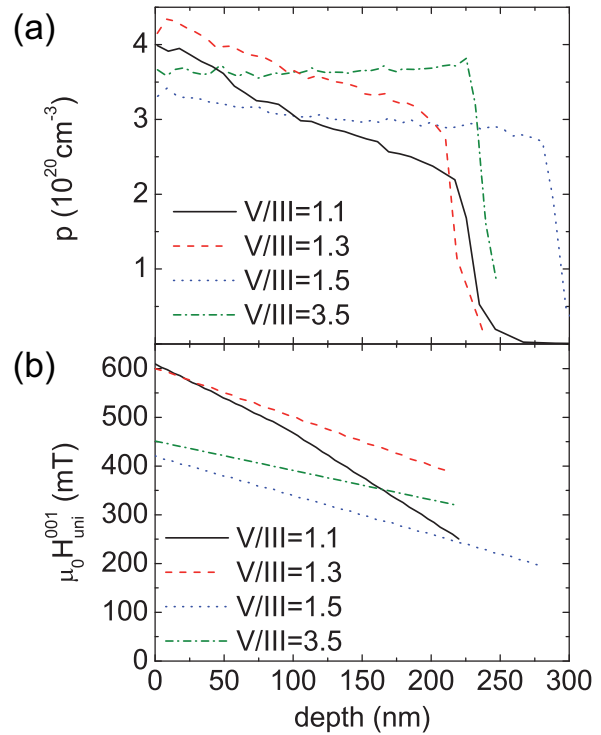}
    \caption{(a) The hole concentration in the different (Ga,Mn)As samples is shown as a function of the depth within the layers as determined by ECV profiling. (b) The uniform resonance fields $\mu_0H^{001}_\text{uni}(z)$ for the four samples obtained from the simulations for the out-of-plane orientation of the external field ($\psi=90^\circ$) as a function of the depth.}
\label{fig:ResonanceFieldGrad_and_ECV}
\end{figure}

(Ga,Mn)As samples with a nominal Mn concentration of $\approx$~4\% were grown on (001)-oriented GaAs substrates by low-temperature molecular-beam epitaxy at a substrate temperature of 220$^\circ$C using V/III flux ratios of 1.1, 1.3, 1.5, and 3.5, referred to as sample A, B, C, and D, respectively. The layer thickness was 210-280~nm as determined from the ECV measurements, cf.~Fig.~\ref{fig:ResonanceFieldGrad_and_ECV}. For samples with V/III flux ratios $\lesssim  3$ a gradient in the hole concentration has been reported,\cite{PRB71_205213} hence this set of samples was chosen to study the influence of a gradient in $p$ on the out-of-plane magnetic anisotropy. Further details on the sample growth can be found in Refs.~\onlinecite{PRB71_205213} and \onlinecite{PRB74_205205}.

The hole concentration profile of the as-grown (Ga,Mn)As layers were determined by ECV profiling using a BioRad PN4400 profiler with a 250~ml aqueous solution of 2.0~g NaOH+9.3~g EDTA as the electrolyte. For further details on the ECV analysis see Ref.~\onlinecite{PRB71_205213}. The results of the ECV measurements for the layers investigated are shown in Fig.~\ref{fig:ResonanceFieldGrad_and_ECV} (a). Except for the sample with V/III=3.5, they reveal a nearly linearly varying hole concentration across the layer thickness with different slopes and with the absolute value of the hole concentration at the surface of the layer varying by about 20\%. The profiles are reproducible within an uncertainty of about 15\%. 
%

\begin{figure*}[h]
\includegraphics[]{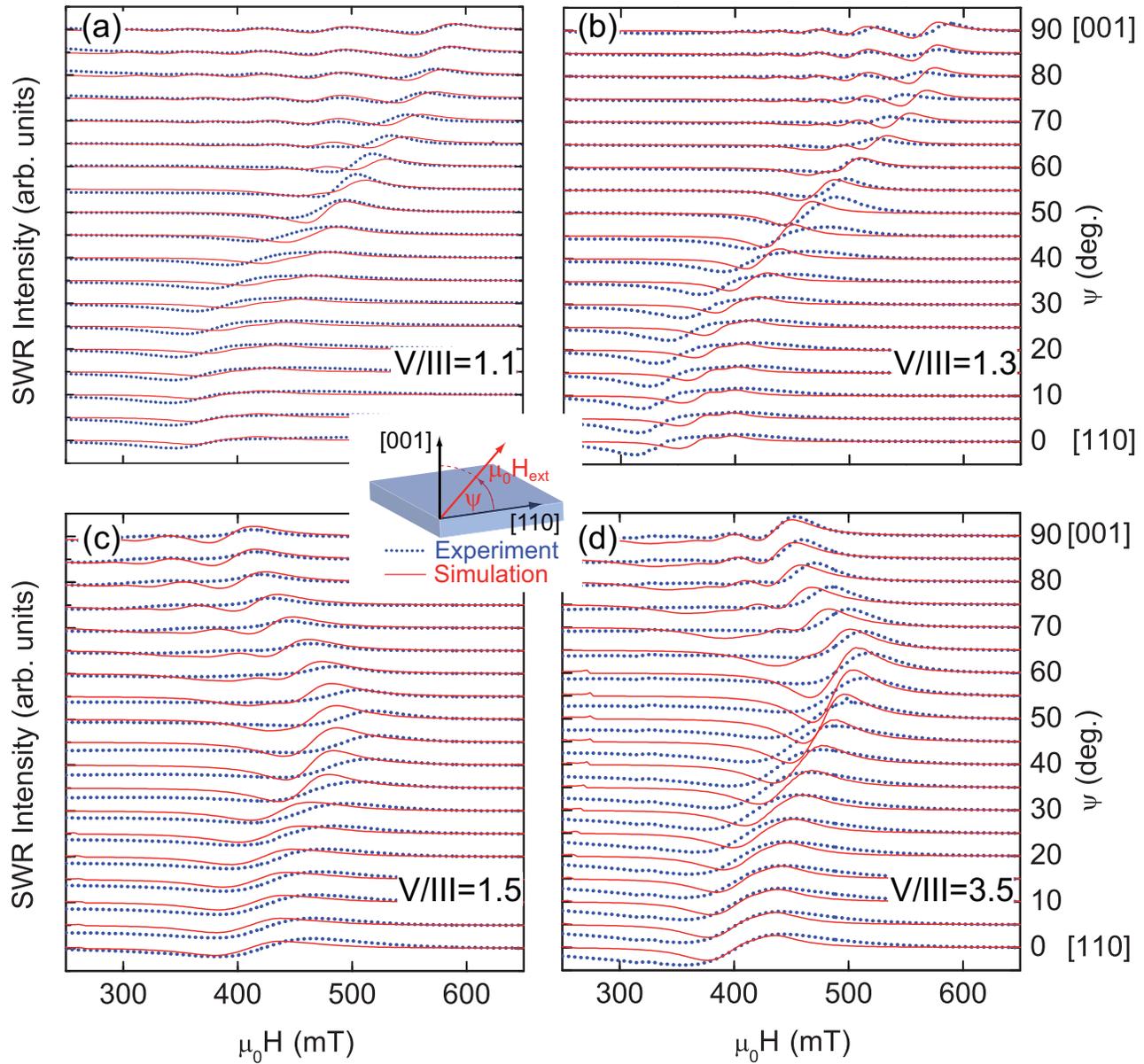}
\centering
\caption{The spin-wave resonance data (dotted, blue lines) are shown together with simulations (red, solid lines) using the numerical procedure described in the text and in the Appendix \ref{subsec:NumericallySWR}. The data were obtained as a function of the external magnetic field orientation and magnitude for samples with a V/III flux ratio of (a) 1.1, (b) 1.3, (c) 1.5, and (d) 3.5. The rotation angle $\psi$ is defined in the inset and the parameters used for the simulations are summarized in Tab.~\ref{tab:AI_parameters}.}
\label{fig:OOPData}
\end{figure*}

To investigate the magnetic anisotropy profiles of the samples, we performed cavity-based FMR measurements, using a Bruker ESP300 spectrometer operating at a microwave frequency of 9.265~GHz ($X$-band) with a microwave power of 2~mW at $T=5~$K; we used magnetic field modulation at a frequency of 100~kHz and an amplitude of 3.2~mT. Since we are mainly interested in the out-of-plane magnetic anisotropy, we recorded spectra for external magnetic field orientations within the crystal plane spanned by the [110] and [001] crystal axes in 5$^\circ$ steps, cf. the inset in Fig.~\ref{fig:OOPData}. For each orientation, the field was ramped to 1~T in order to saturate the magnetization and then swept from 650~mT to 250~mT; the spectra for the samples investigated are shown in Fig.~\ref{fig:OOPData}. 

We start by discussing qualitative differences in the spectra. The samples A and B exhibit several pronounced resonances for the external field oriented along [001], which we attribute to standing spin-wave resonances [Fig.~\ref{fig:OOPData} (a) and (b)]. For these samples, the [001] direction is the magnetically hardest axis since at this orientation the resonance field of the fundamental spin-wave mode is larger than at all other orientations. As the external field is rotated into the film plane, the resonance position of this mode gradually shifts to lower field values as expected for a pronounced out-of-plane hard axis. In contrast, the samples C and D exhibit the largest resonance fields for a field orientation of 50-60$^\circ$ [Fig.~\ref{fig:OOPData} (c) and (d)] pointing to an interplay of second- and fourth-order out-of-plane anisotropy with different signs of the corresponding anisotropy parameters. These samples exhibit spin-wave resonances as well, however they are less pronounced than for samples A and B.

To quantitatively model the spin-wave spectra we numerically solve for each magnetic field orientation the spin-wave equation \eqref{eq:SpinWaveEquationElliptical} by the finite difference method as outlined in the Appendix \ref{subsec:NumericallySWR}. Although this method allows for the modeling of the SWR for arbitrary profiles of the anisotropy parameters, the exchange stiffness, the Gilbert damping parameter, and the saturation magnetization, we assume the parameters to vary linearly as a function of $z$. This approach is motivated by the linear gradient in the hole concentration, which in first approximation is assumed to cause a linear gradient in the anisotropy parameters, resulting in the spin-wave resonances observed in the samples.\cite{APL82_730,PRB79_45205} In Tab.~\ref{tab:AI_parameters}, we have summarized the parameters used in the simulation for the different samples. The parameters in capital letters denote the value at the surface of the sample while the ones in lower-case letters denote the slope of this parameter; e.g., the $z$ dependence of the second-order, uniaxial out-of-plane anisotropy parameter is given by $B_\mathrm{001}(z)= B_\mathrm{001}-b_\mathrm{001}\times z$. The layer thickness used for the simulation can be inferred from Fig.~\ref{fig:ResonanceFieldGrad_and_ECV} (a) and was determined from the ECV data under the assumption that at the position where the hole concentration rapidly decreases the magnetic properties of the layer abruptly undergo a transition from ferromagnetic to paramagnetic.
For the simulations, we divided each film into $n=100$ layers with constant magnetic properties within each layer. For the gyromagnetic ratio we used $\gamma=g\mu_\text{B}/\hbar$ with $g=2$.\cite{APL89_012507}

As result of the simulation we obtain the Polder susceptibility tensor $\bar{\chi}(\mu_0 \boldsymbol{H},z)$ and the transverse magnetization components as a function of $z$ and $\mu_0\boldsymbol{H}$. Additionally, we obtain the $z$ dependence of the uniform resonance field by solving Eq.~\eqref{eq:UniformResonanceField} for each field orientation. In an SWR absorption experiment with magnetic field modulation, the obtained signal is proportional to the first derivative of the absorbed microwave power with respect to the magnetic field. Thus, we calculate the absorbed power using the simulated susceptibility and Eq.~\eqref{eq:absorbedPower} and numerically differentiate the result in order to compare the simulated SWR spectra with the experiment. Additionally, we use a global scaling factor, accounting, e.g., for the modulation amplitude, which is the same for all field orientations, and we multiply all the simulated data with this factor. In Fig.~\ref{fig:OOPData}, we plot the experimental data together with the simulations using the parameters given in Tab.~\ref{tab:AI_parameters}, demonstrating that a reasonable agreement between theory and experiment can be found with one set of simulation parameters for all magnetic field orientations for each sample. 
\begin{table*}[htp]
\centering
\caption{Simulation parameters and their $z$ dependence of the samples under study as obtained by fitting the simulations to the SWR measurements. For the anisotropy parameters the capital letters denote the value at the surface of the film and the lower case letters the slope as described in the text. For sample A, the first value of $b_{001}$ was used for the first 100~nm and the second one for the remaining layer. In addition to the anisotropy parameters, the saturation magnetization is also assumed to
vary linearly across the layer, while its absolute value is unknown and not important for the SWR simulations.}
\begin{ruledtabular}
\begin{tabular}{c c c c c c c c c c c}
Sample &V/III & $B_{001}$ & $b_{001}$ & $B_{4\parallel}$ & $b_{4\parallel}$ & $B_{4\perp}$ & $b_{4\perp}$ & $D_\text{s}$ & $\alpha$ & $\frac{\partial M(z)}{\partial z M(0)}$ \\\hline
 && (mT) & ($\frac{\text{mT}}{\text{nm}}$) & (mT) & ($\frac{\text{mT}}{\text{nm}}$) & (mT) & ($\frac{\text{mT}}{\text{nm}}$) & (Tnm$^2$) & & ($\frac{1}{\mu\text{m}}$)\\\hline
A & 1.1 &90&-0.1, -0.3&-50&0.05&25&-0.3& 35 & 0.09 & -3 \\
B & 1.3 &130&-0.5&-50&0&0 &0 & 20 & 0.06 & -4\\
C & 1.5 &75&-0.4&-55&-0.04&-15 & 0& 40 & 0.11 & -4\\
D & 3.5 &91&-0.3&-55&-0.04&-15& 0 & 20 & 0.09 & -3
\label{tab:AI_parameters}
\end{tabular}
\end{ruledtabular}
\end{table*}

\begin{figure*}[h]
\includegraphics[]{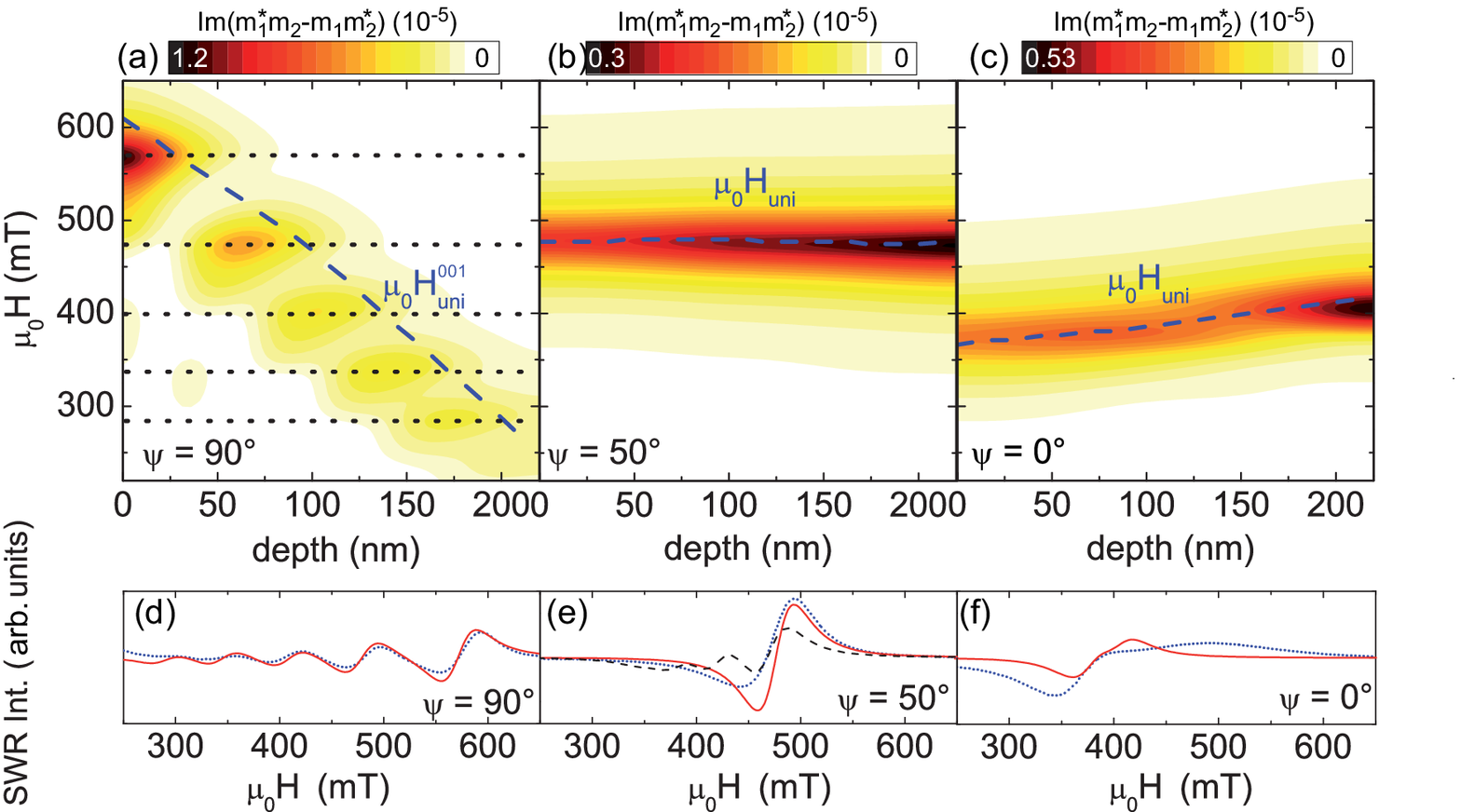}
\caption{Simulated magnetization mode profile and uniform resonance field of sample A. The contour plots show the magnetization precession amplitude $\text{Im}(m_1^*m_2-m_1m_2^*)$ as a function of the position within the film and the external magnetic field for the external field aligned (a) along [001], (b) at an angle of 50$^\circ$ with respect to [110] (cf.~the inset in Fig.~\ref{fig:OOPData}) and (c) along [110]. The blue, dashed lines in (a)-(c) show the uniform resonance field, obtained by numerically solving Eq.~\eqref{eq:UniformResonanceField} for each given field orientation. The dotted black lines in (a) indicate the resonance magnetic fields. In (d)-(f), a magnification of the data (blue dotted lines) and simulation (red solid lines) from Fig.~\ref{fig:OOPData} (a) is shown using the same scale for all orientations. In (e), a simulation with a different set of parameters is shown for comparison (black, dashed line), see text.}
\label{fig:MagnetizationProfiles}
\end{figure*}%

We will now exemplarily discuss the angle dependence of the SWR spectrum of sample A shown in Fig.~\ref{fig:OOPData}(a) based on the uniform resonance field and the resulting magnetization mode profile obtained from the simulation. To this end, we plot in Fig.~\ref{fig:MagnetizationProfiles} (a)-(c) the magnetization precession amplitude $\text{Im}(m_1^*m_2-m_1m_2^*)$ for selected external field orientations as a function of depth and external magnetic field in a contour plot, together with the corresponding uniform resonance field. In Fig.~\ref{fig:MagnetizationProfiles} (d)-(f), we show for each external field orientation a magnification of the corresponding SWR spectrum together with the simulation. Note that in contrast to the normal-mode approach (Appendix \ref{subsec:NumericallySWR_simple}) used to calculate the modes in Fig.~\ref{fig:DidaktikSWR}, where the coupling of each mode to the cavity field has to be found by integration, the approach elaborated in the Appendix \ref{subsec:NumericallySWR} directly yields the transverse magnetization components, already accounting for the coupling efficiency and the linewidth. Further, the approach presented in the Appendix \ref{subsec:NumericallySWR} is also valid when the difference in the resonance fields of two modes is comparable with or smaller than their linewidth, in contrast to the normal-mode approach \cite{Hoekstra_diss}.

If the external field is parallel to the surface normal ($\psi=90^\circ$) the uniform resonance field varies by about 350~mT across the film thickness [cf.~the dashed line in Fig.~\ref{fig:MagnetizationProfiles} (a)], resulting in several well-resolved standing spin-wave modes. The spin-wave resonance fields are plotted as dotted lines in Fig.~\ref{fig:MagnetizationProfiles} (a); since the spacing of the resonance fields is larger than the SWR linewidth, the modes are clearly resolved, cf.~Fig.~\ref{fig:MagnetizationProfiles} (a) and (d). In the simulation two regions with different $b_{001}$ values were used in order to reproduce the spacing of the higher-order spin-wave modes found in the experiment. Using the same slope as in the first 100~nm for the entire layer would lead to a smaller spacing between the third and higher order modes. Instead of defining two regions with different slopes $b_{001}$, a gradient in the exchange stiffness with positive slope could also be used to model the experimentally found mode spacing as discussed in the context of Fig.~\ref{fig:Bildatlas}. Since the exchange interaction in (Ga,Mn)As is mediated by holes\cite{PRB63_195205} and $p$ decreases across the layer, we refrain from modeling our results with a positive gradient in $D_\text{s}$. Further, the results in Ref.~\onlinecite{PRB79_45205} rather point to a negative gradient in $D_\text{s}$ in a similar sample. However, a decreasing Mn concentration as a function of the depth could lead to an increase of $D_\text{s}$.\cite{MITo43_3019}

Finally, we note, since $B_{1\bar{1}0}=0$ in the simulations, the magnetization precesses circularly for $\psi=90^\circ$ and thus $\text{Im}(m_1^*m_2-m_1m_2^*)=2\sin^2{\tau}$,\cite{PRB86_134415} with the precession cone angle $\tau$. For all other orientations, $\vec{m}$ precesses elliptically which is accounted for in our simulations. In the simulations of the precession amplitudes, we have assumed an externally applied microwave magnetic field with $\mu_0h=0.1$~mT.

At an external field orientation of $\psi=50^\circ$ the uniform resonance field is nearly constant across the layer, and consequently only one SWR mode is observed with an almost uniform magnetization precession across the layer, cf.~Fig.~\ref{fig:MagnetizationProfiles} (b). The precession amplitude is a measure for the SWR intensity. While the fundamental mode at $\psi=90^\circ$ exhibits a larger precession cone at the interface, it rapidly decays as a function of the depth, in contrast to the nearly uniform precession amplitude for $\psi=50^\circ$. Since the entire layer contributes to the power absorption, consequently, the SWR mode at $\psi=50^\circ$ is more intense than the fundamental mode for $\psi=90^\circ$, which is indeed observed in the experiment [cf.~Fig.~\ref{fig:MagnetizationProfiles} (d) and (e)].

For the magnetic field within the film plane [$\psi=0^\circ$, cf.~Fig.~\ref{fig:MagnetizationProfiles} (c)], the uniform resonance field again varies linearly across the film, however in a less pronounced way than for the out-of-plane field orientation and with an opposite sign of the slope. The sign reversal of the slope can be understood in terms of the uniaxial out-of-plane anisotropy parameter $B_{001}$: positive values of these parameters lead to an increase (decrease) of the resonance field for the magnetization oriented perpendicular (parallel) to the film plane, accounting for the slopes of the uniform resonance fields in Fig.~\ref{fig:MagnetizationProfiles}.
Since the gradient in the uniform resonance field is less pronounced for $\psi=0^\circ$ than for $\psi=90^\circ$, the spin-wave modes are not resolved for $\psi=0^\circ$, since their spacing is smaller than the SWR linewidth, leading to one rather broad line [cf.~Fig.~\ref{fig:MagnetizationProfiles} (c) and (f)]. A steeper gradient in $B_{4||}$ in combination with a different Gilbert damping (or with an additional inhomogeneous damping parameter) and amplitude scaling factor, could improve the agreement of simulation and experiment in the in-plane configuration, as discussed later. A detailed study of the in-plane anisotropy profile is however beyond the scope of this work. Given that the presented simulations were obtained with one set of parameters, the agreement of theory and experiment is reasonably good also for the in-plane configuration, since salient features of the SWR lineshape are reproduced in the simulation.

Having discussed the angle-dependence of the SWR spectra, we turn to the $z$ dependence of the out-of-plane anisotropy of sample A. Our simulations reveal that it is governed by the $z$ dependence of both $B_{001}(z)$ and $B_{4\perp}(z)$. Assuming only a gradient in $B_{001}$ results in a reasonable agreement of theory and experiment for the external field oriented along [001] and [110], but fails to reproduce the spectra observed for the intermediate field orientations, e.g., $\psi=50^\circ$. This is illustrated by the dashed black line in Fig.~\ref{fig:MagnetizationProfiles} (e), which represent simulations with a constant $B_{4\perp}(z)$ for $\psi=50^\circ$. As can be seen, this simulation produces several spin-wave resonances, whereas in the experiment only one resonance is present, which is better reproduced by the simulation with both $B_{001}(z)$ and $B_{4\perp}(z)$ varying across the layer.

We will now discuss the anisotropy parameters of all samples. In contrast to sample A, the out-of-plane anisotropy profile of all other samples appears to be governed by a gradient in $B_{001}(z)$. As already discussed qualitatively, the hard axis of the samples is determined by an interplay of $B_{001}$ and $B_{4\perp}$. For sample A and B $B_{4\perp}$ is positive and zero, leading to an out-of-plane hard axis. In contrast, sample C and D exhibit a negative $B_{4\perp}$, leading to a hard axis between out-of-plane and in-plane. The $B_{4||}$ parameter is negative and of similar magnitude for all samples. 

%

Since the out-of-plane anisotropy profile of sample A is governed by $B_{001}(z)$ and $B_{4\perp}(z)$, a comparison of the out-of-plane anisotropy profile between all samples based on anisotropy parameters is difficult. We therefore compare the uniform resonance fields, where both anisotropy parameters enter. As evident from Fig.~\ref{fig:MagnetizationProfiles}, the strongest influence of the magnetic inhomogeneity of the layers on the uniform resonance fields is observed for the external field along [001]. To compare the hole concentration profile in Fig.~\ref{fig:ResonanceFieldGrad_and_ECV} (a) with the anisotropy profile, we therefore plot in Fig.~\ref{fig:ResonanceFieldGrad_and_ECV} (b) the $z$ dependence of the uniform resonance field $\mu_0\boldsymbol{H}^{001}_\text{uni}$ for this field orientation. The figure demonstrates that the gradient in $\mu_0\boldsymbol{H}^{001}_\text{uni}$ is correlated with the gradient in $p$. For the sample with the strongest gradient in $p$ the gradient in $\mu_0\boldsymbol{H}^{001}_\text{uni}$ is also most distinct while the samples with a weaker gradient in $p$ exhibit a less pronounced gradient in $\mu_0\boldsymbol{H}^{001}_\text{uni}$. However, for sample D, exhibiting a nearly constant $p$, we still observe standing spin wave resonances for $\mu_0\boldsymbol{H}||[001]$ [Fig.~\ref{fig:OOPData} (d)], reflected in a slight gradient of $\mu_0\boldsymbol{H}^{001}_\text{uni}$. This observation suggests that aditionally other mechanisms lead to a variation of the anisotropy profile. One possibility would be a gradient in the elastic strain of the layer, due to a non-homogeneous incorporation of Mn atoms in the lattice. However, x-ray diffraction measurements of this sample, in combination with a numerical simulation based on dynamic scattering theory, reveal a variation of the vertical strain $\Delta\varepsilon_{zz}$ as small as $3\times 10^{-5}$ across the layer. According to the measurements in Ref.~\onlinecite{PRB79_195206}, such a variation in strain would lead to a variation of the $B_{001}$ parameter by a few mT only, insufficient to account for the variation of $\mu_0\boldsymbol{H}_\text{uni}$ by almost 100~mT across the layer. A more likely explanation seems to be a variation of the saturation magnetization, which should also influence the anisotropy parameters. In the simulation, a non-homogeneous saturation was assumed, potentially explaining also the observed gradient in the anisotropy parameters and therefore in the uniform resonance field.

In contrast to the out-of-plane anisotropy parameters, $B_{4||}$ was found to depend only weakly on $z$, for all samples except sample B where it was constant. Additionally, $B_{1\bar{1}0}$, typically of the order of a few mT,\cite{PRB79_195206} might have an influence and interplay with $B_{4||}$ in determining the in-plane anisotropy. We here however focus on the out-of-plane anisotropy and therefore neglect $B_{1\bar{1}0}$ in our simulations. An in-plane rotation of the external field would be required for a more accurate measurement of $B_{4||}$ and $B_{1\bar{1}0}$, but is outside the scope of this work.

According to the valence-band model in Ref.~\onlinecite{PRB63_195205}, an oscillatory behavior of the magnetic anisotropy parameters is expected as a function of $p$. Therefore, depending on the absolute value of $p$, different values for, e.g., $\partial B_{001}/\partial p$ are expected. In particular, there are regions where a anisotropy parameter might be nearly independent of $p$ and other regions with a very steep $p$ dependence. Since the absolute value of $p$ is unknown, a quantitative discussion of the $p$ dependence of the obtained anisotropy parameters based on the model in Ref.~\onlinecite{PRB63_195205} is not possible. 
In addition to $p$, the $p$-$d$ exchange integral,\cite{PRB63_195205} which may also vary as a function of the depth in a non-homogneous film, also influences the anisotropy parameters,\cite{PRB63_195205} further complicating a quantitative analysis.

For all samples, we used a constant exchange stiffness $D_\text{s}$ in our modeling. As alluded to above, there is some ambiguity in this assumption, since the exchange stiffness and the gradient in the anisotropy both influence the mode spacing. For simplicity, however, we intended to keep as many simulation parameters as possible constant. The absolute values obtained for the exchange stiffness agree within a factor of 2 with the ones obtained in previous experiments\cite{PRB75_233308,PRB79_45205} but are a factor of 2-4 larger than theoretically predicted.\cite{PRB82_85204} For the reasons discussed above, there is a large uncertainty also in the derivation of the absolute value of $D_\text{s}$ from standing spin-wave modes in layers with a gradient in the magnetic anisotropy constants. 

In order to use one parameter set for all field-orientations, the Gilbert damping parameter was assumed to be isotropic in the simulations. The modeling of the SWR data could be further improved by assuming a non-isotropic damping, its value being larger for $\mu_0\boldsymbol{H}||[110]$ than for $\mu_0\boldsymbol{H}||[001]$ [cf.~Fig.~\ref{fig:OOPData}]. This however, only improves the result when assuming a field orientation-dependent scaling factor for the amplitude, which could be motivated, e.g., by the assumption that the microwave magnetic field present at the sample position depends on the sample orientation within the cavity. The absolute values of $\alpha$ determined here are comparable with the ones obtained by ultra-fast optical experiments,\cite{APL91_3} but are larger than the typical $\alpha=0.01...0.03$ values found by frequency-dependent FMR studies.\cite{APL92_3,PRB78_195210} As already alluded to, inhomogeneous line-broadening mechanisms may play a dominant role,\cite{PRB78_195210} in particular for as-grown samples.\cite{PRB69_85209} We therefore assume that the values for $\alpha$ obtained in this study overestimate the actual intrinsic Gilbert damping. A frequency-dependent SWR study would be required to determine the intrinsic $\alpha$. Such a study could possibly also reveal a $p$-dependent $\alpha$ as theoretically predicted.\cite{PRB69_85209} In our study, assuming a $z$ dependent $\alpha$ did not improve the agreement between simulation and experiment, corroborating the conjecture that inhomogeneous broadening mechanisms dominate the linewidth and therefore obscure a possible $z$ dependence of $\alpha$.

\section{Summary}\label{summary}
We have presented a finite difference-type modeling approach for standing spin-wave resonances based on a numerical solution of the LLG equation. With this generic formalism, SWR spectra can be simulated accounting for elliptical magnetization precession, for arbitrary orientations of the external magnetic field, and for arbitrary profiles of all magnetic properties, including anisotropy parameters, exchange stiffness, Gilbert damping, and saturation magnetization. The approach is applicable not only to (Ga,Mn)As but to all ferromagnets. 

Four (Ga,Mn)As samples, epitaxially grown with V/III flux ratios of 1.1, 1.3, 1.5, and 3.5 were investigated by ECV and spin-wave resonance spectroscopy, revealing a correlation of a linear gradient in the hole concentration with the occurrence of standing spin wave resonances, in particular for the external field oriented out-of-plane. Using the presented modeling approach, the SWR spectra could be reproduced in a simulation with one parameter set for all external field orientations. The simulation results demonstrate that the profile of the out-of-plane uniform resonance field is correlated with the hole concentration profile. However, our measurements and simulations show, that a non-uniform hole concentration profile is not the only cause that leads to the observed non-uniform magnetic anisotropy; possibly, a variation in the saturation magnetization also influences the anisotropy parameters. To gain a quantitative understanding of this issue, more samples with known hole concentrations would be required, where both the absolute values and the profiles of $p$ are varied. Such a study was, however, outside the scope of this work.

Besides the modeling of SWR intensities and linewidths, the presented formalism yields the magnetization precession amplitude as a function of the position within the ferromagnet. It can therefore be used to investigate spin-pumping intensities in (Ga,Mn)As/Pt bilayers.\cite{PRL107_46601} The spin-pumping signal, detected as a voltage across the Pt layer, should be proportional to the magnetization precession cone in the vicinity of the (Ga,Mn)As/Pt interface. By measuring the spin-pumping signal as well as the SWR intensities of (Ga,Mn)As/Pt and by using our modeling approach, it should be possible to investigate to which extent a magnetization mode which is localized at a certain position within the (Ga,Mn)As layer contributes to the spin-pumping signal.

\begin{acknowledgments}
This work was supported by the Deutsche Forschungs-Gemeinschaft via Grant No. SFB 631 C3  (Walter Schottky Institut) and Grant No. Li 988/4 (Universit\"at Ulm).
\end{acknowledgments}

\appendix

\section{Coordinate Transformation and Free Enthalpy derivatives\label{Appendix_A_SWR}}
The transformation between the crystallographic coordinate system ($x,y,z$) and the equilibrium system (1,2,3) is given by 
\begin{equation}
     \left(\begin{array}{c}
     m_x \\
     m_y\\
     m_z\\
     \end{array}\right)=T \left(\begin{array}{c}
     m_1 \\
     m_2\\
     m_3\\
     \end{array}\right),
    \label{eq:SWR_Transformation}
\end{equation}
with
\begin{equation}
T=\left(\begin{array}{ccc}
     \cos\theta_0\cos\phi_0 & -\sin\phi_0 & \sin\theta_0\cos\phi_0 \\
     \cos\theta_0\sin\phi_0 & \cos\phi_0 &\sin\theta_0\sin\phi_0\\
     -\sin\theta_0 & 0 & \cos\theta_0 \\
     \end{array}\right).
    \label{eq:SWR_TransformationMatrix}
\end{equation}
The derivatives of the free enthalpy density Eq.~\eqref{eq:SWR_Expansion_G} with respect to the magnetization components are
\begin{eqnarray}
G_3&=&\partial_{m_3}G|_{\vec{m}=\vec{m_0}}=-\mu_0H_3+2B_{001}\cos^2\theta_0\nonumber\\
	 &+&B_{1\bar{1}0}(\sin\theta_0\cos\phi_0-\sin\theta_0\sin\phi_0)^2\nonumber\\
	 &+&4B_{4\perp}\cos^4\theta_0 \nonumber\\
	 &+&4B_{4\parallel}\sin^4\theta_0(\cos^4\phi_0+\sin^4\phi_0)\\
\label{eq:SWR_G_3}
G_{21}&=&G_{12}=\partial_{m_1}\partial_{m_2}G|_{\vec{m}=\vec{m_0}}\nonumber\\
&=&\cos\theta_0(1-2\cos^2\phi_0)[B_{1\bar{1}0}\nonumber\\
&+&12B_{4\parallel}\sin^2\theta_0\cos\phi_0\sin\phi_0]\label{eq:SWR_G_12}\\
G_{11}&=&\partial_{m_1}\partial_{m_1}G|_{\vec{m}=\vec{m_0}}=2B_{001}\sin^2\theta_0\nonumber\\
&+&12\cos^2\theta_0\sin^2\theta_0[B_{4\perp}\nonumber\\
&+&B_{4\parallel}(\cos^4\phi_0+\sin^4\phi_0)]\nonumber\\
   &+&B_{1\bar{1}0}\cos^2\theta_0(\cos\phi_0-\sin\phi_0)^2 \label{eq:SWR_G_11}\\
G_{22}&=&\partial_{m_2}\partial_{m_2}G|_{\vec{m}=\vec{m_0}}=2B_{1\bar{1}0} (\sin\phi_0+\cos\phi_0)^2\nonumber\\
   &+&24 B_{4\parallel}\sin^2\theta_0\cos^2\phi_0\sin^2\phi_0.\label{eq:SWR_G_22}
\end{eqnarray}

\section{Finite Difference Method\label{Appendix_B_SWR}}
In this Appendix, we describe how the spin-wave equation can be numerically solved by the finite difference method. We start with the simple case of a circulary precessing magnetization, neglecting Gilbert damping and the driving field (Sec.~\ref{subsec:NumericallySWR_simple}). Then we turn to the general case, where the magnetization precesses elliptically and the Gilbert damping as well as the driving field are included (Sec.~\ref{subsec:NumericallySWR}).

\subsection{The One-Dimensional, Homogeneous, Undamped Case}\label{subsec:NumericallySWR_simple}
Here, we describe how the resonance fields and the spin-wave modes can be found, assuming a circularly precessing magnetization $m_2=i m_1=\tilde{m}$, a constant exchange stiffness, and a $z$ independent equilibrium magnetization. This case has been considered in Ref.~\citenum{PRB79_45205} using a semi-analytical approach to solve the spin-wave equation Eq.~\eqref{eq:SpinWaveEquationSimple}. The approach considered here, is slightly more general, as it is straight forward to determine resonance fields and eigenmodes of the system for an arbitrary $z$ dependence of the uniform resonance field.
To solve Eq.~\eqref{eq:SpinWaveEquationSimple}, we divide the ferromagnetic film into a finite number $n$ of layers with equal thickness $l$ and constant magnetic properties within each of these layers. The $z$ dependence of $\tilde{m}$ and $\mu_0H^{001}_{uni}$ is thus given by an index $j=1...n$. Within each of these layers the uniform resonance field and $\tilde{m}(z)$ are thus constant and given by the values $\mu_0H^{001,j}_{uni}=:K^j$ and $\tilde{m}^j$, respectively. The second derivative of $\tilde{m}$ is approximated by
\begin{equation}
\tilde{m}''(z=j\cdot l)\approx\frac{\tilde{m}^{j-1}-2\tilde{m}^{j}+\tilde{m}^{j+1}}{l^2}.
\label{eq:2ndDerivative_mtilde}
\end{equation}
Consequently, Eq.~\eqref{eq:SpinWaveEquationSimple} is converted to the homogeneous equation system
\begin{widetext}
\begin{equation}
\left(\begin{array}{cccccccc}
   & \vdots &\vdots &\vdots & \\
   \hdots &K^{j-1}\!+\!2d & -d & 0 & \hdots\\
   \hdots &-d & K^{j}\!+\!2d & -d & \hdots\\
   \hdots &0 & -d & K^{j+1}\!+\!2d& \hdots\\
   & \vdots &\vdots &\vdots & \\
   \end{array}\right)
\left(\begin{array}{c}
   \vdots \\
   \tilde{m}^{j-1}\\
   \tilde{m}^{j}\\
   \tilde{m}^{j+1}\\
   \vdots\\
   \end{array}\right)
   =\mu_0H\left(\begin{array}{c}
   \vdots \\
   \tilde{m}^{j-1}\\
   \tilde{m}^{j}\\
   \tilde{m}^{j+1}\\
   \vdots\\
   \end{array}\right),
\label{eq:SpinWaveSimpleNumerically}
\end{equation}
\end{widetext}
with the abbreviation $d=-D_\text{s}/l^2$. The boundary condition of natural freedom \cite{PRB79_45205} (von Neumann boundary condition) reads as $\tilde{m}^0=\tilde{m}^1$ and $\tilde{m}^{n-1}=\tilde{m}^n$ and can be incorporated in Eq.~\eqref{eq:SpinWaveSimpleNumerically}. Since the matrix on the left hand side of Eq.~\eqref{eq:SpinWaveSimpleNumerically} is sparse, it can be efficiently diagonalized numerically, yielding the resonance fields (eigenvalues) and the corresponding modes (eigenvectors). After diagonalizing the matrix, the relevant resonance fields are found by sorting the eigenvalues and considering only the modes with positive resonance fields, corresponding to the bound states in the particle-in-a-box analogon.
The SWR amplitude of each mode is proportional to its net magnetic moment; thus, the amplitudes can be found by integrating the (normalized) eigenmodes. The mode profile, the resonance fields, and the SWR intensities are illustrated in Fig.~\ref{fig:DidaktikSWR} for a constant and a linearly varying uniform resonance field. The finite linewidth of the SWR modes can be accounted for by assuming a Lorentzian lineshape for each mode with a certain linewidth and with the resonance fields and intensities calculated as described above \cite{PRB79_45205}. Note that this approach to derive resonance fields and intensities is only valid if the mode separation is large compared with the linewidth of the modes; this restriction does not apply to the model presented in the Appendix \ref{subsec:NumericallySWR}.

\subsection{The General Case}\label{subsec:NumericallySWR}
To solve Eq.~\eqref{eq:SpinWaveEquationElliptical} for arbitrary $\mu_0\boldsymbol{H}$ and arbitrarily varying magnetic properties, we again divide the ferromagnetic film into a finite number $n$ of layers with equal thickness $l$ and constant magnetic properties within each of these layers. In contrast to the case in the Appendix \ref{subsec:NumericallySWR_simple}, where only the uniform resonance field was varied across the layer, here potentially all magnetic properties entering Eq.~\eqref{eq:SpinWaveEquationElliptical} can be assumed to be $z$ dependent. Additionally, the components of the driving field $\mu_0h_i$ ($i=1,2$), can also vary as a function of $z$, since the $(1,2,3)$ frame of reference is $z$ dependent and thus the projections of the driving field have to be calculated for each layer. The $z$ dependence of the components $m_i$ ($i=1,2$), of the parameters $H_{11}$, $H_{12}$, $H_{21}$, $H_{22}$ (defined in Sec.~\ref{theory}) and the exchange stiffness is thus given by the index $j=0...n$; the second derivative of each of the components $m_i$ is approximated as in Eq.~\eqref{eq:2ndDerivative_mtilde}.


The linearized LLG equation Eq.~\eqref{eq:SpinWaveEquationElliptical}, is thus converted into the inhomogeneous equation system
%
%
\begin{widetext}
\begin{equation}
\setlength{\arraycolsep}{0.0em}
\left(\begin{array}{cccccccc}
   & \vdots &\vdots &\vdots &\vdots &\vdots &\vdots & \\
   \hdots &H_{11}^{j-1}\!-\!2d^{j-1} & H_{12}^{j-1} & d^{j-1} & 0 & 0 & 0 & \hdots\\
   \hdots &H_{21}^{j-1} & H_{22}^{j-1}\!-\!2d^{j-1} & 0 & d^{j-1} & 0 & 0 & \hdots\\
   \hdots &d^j & 0 & H_{11}^{j}\!-\!2d^{j} & H_{12}^{j} & d^{j} & 0 & \hdots\\
   \hdots & 0 &d^{j} & H_{21}^{j} & H_{22}^{j}\!-\!2d^{j} & 0 & d^{j} & \hdots\\
   \hdots &0 & 0 &d^{j+1} & 0 & H_{11}^{j+1}\!-\!2d^{j+1} & H_{12}^{j+1} & \hdots\\
   \hdots &0 & 0 &0 & d^{j+1} & H_{21}^{j+1} & H_{22}^{j+1}\!-\!2d^{j+1} & \hdots\\
   & \vdots &\vdots &\vdots &\vdots &\vdots &\vdots & \\
   \end{array}\right)
\left(\begin{array}{c}
   \vdots \\
   m_1^{j-1}\\
   m_2^{j-1}\\
   m_1^{j}\\
   m_2^{j}\\
   m_1^{j+1}\\
   m_2^{j+1}\\
   \vdots\\
   \end{array}\right)
   =\mu_0\left(\begin{array}{c}
   \vdots \\
   h_{1}^{j-1}\\
   h_{2}^{j-1}\\
   h_{1}^{j}\\
   h_{2}^{j}\\
   h_{1}^{j+1}\\
   h_{2}^{j+1}\\
   \vdots\\
   \end{array}\right),
\label{eq:SpinWaveNumerically}
\end{equation}
\end{widetext}
%
with the abbreviation $d^j=-D_\text{s}^j/l^2$.
At the boundaries of the magnetic film we again assume the spins to exhibit natural freedom $m_i^0=m_i^1$ and $m_i^n=m_i^{n+1}$.

To simulate a spin-wave spectrum for a given orientation of the external field and a given profile of the magnetic properties, we numerically sweep the magnetic field and calculate the equilibrium magnetization orientation for all indices $j=0...n$ at a given external field. The inverse of the matrix in Eq.~\eqref{eq:SpinWaveNumerically}, multiplied by $\mu_0 M(z)$, is the generalized Polder susceptibility tensor $\bar{\chi}(\mu_0 \boldsymbol{H},z)$, which relates the transverse magnetization with the driving field, cf.~Eq.~\eqref{eq:Polder}.



\begin{thebibliography}{55}
\expandafter\ifx\csname natexlab\endcsname\relax\def\natexlab#1{#1}\fi
\expandafter\ifx\csname bibnamefont\endcsname\relax
  \def\bibnamefont#1{#1}\fi
\expandafter\ifx\csname bibfnamefont\endcsname\relax
  \def\bibfnamefont#1{#1}\fi
\expandafter\ifx\csname citenamefont\endcsname\relax
  \def\citenamefont#1{#1}\fi
\expandafter\ifx\csname url\endcsname\relax
  \def\url#1{\texttt{#1}}\fi
\expandafter\ifx\csname urlprefix\endcsname\relax\def\urlprefix{URL }\fi
\providecommand{\bibinfo}[2]{#2}
\providecommand{\eprint}[2][]{\url{#2}}

\bibitem[{\citenamefont{Zemen et~al.}(2009)\citenamefont{Zemen, Kucera,
  Olejnik, and Jungwirth}}]{PRB80_155203}
\bibinfo{author}{\bibfnamefont{J.}~\bibnamefont{Zemen}},
  \bibinfo{author}{\bibfnamefont{J.}~\bibnamefont{Kucera}},
  \bibinfo{author}{\bibfnamefont{K.}~\bibnamefont{Olejnik}}, \bibnamefont{and}
  \bibinfo{author}{\bibfnamefont{T.}~\bibnamefont{Jungwirth}},
  \bibinfo{journal}{Phys. Rev. B} \textbf{\bibinfo{volume}{80}},
  \bibinfo{pages}{155203} (\bibinfo{year}{2009}).

\bibitem[{\citenamefont{Liu et~al.}(2005)\citenamefont{Liu, Lim, Ge, Shen,
  Dobrowolska, Furdyna, Wojtowicz, Yu, and Walukiewicz}}]{APL86_112512}
\bibinfo{author}{\bibfnamefont{X.}~\bibnamefont{Liu}},
  \bibinfo{author}{\bibfnamefont{W.~L.} \bibnamefont{Lim}},
  \bibinfo{author}{\bibfnamefont{Z.}~\bibnamefont{Ge}},
  \bibinfo{author}{\bibfnamefont{S.}~\bibnamefont{Shen}},
  \bibinfo{author}{\bibfnamefont{M.}~\bibnamefont{Dobrowolska}},
  \bibinfo{author}{\bibfnamefont{J.~K.} \bibnamefont{Furdyna}},
  \bibinfo{author}{\bibfnamefont{T.}~\bibnamefont{Wojtowicz}},
  \bibinfo{author}{\bibfnamefont{K.~M.} \bibnamefont{Yu}}, \bibnamefont{and}
  \bibinfo{author}{\bibfnamefont{W.}~\bibnamefont{Walukiewicz}},
  \bibinfo{journal}{Appl. Phys. Lett.} \textbf{\bibinfo{volume}{86}},
  \bibinfo{pages}{112512} (\bibinfo{year}{2005}).

\bibitem[{\citenamefont{Dreher et~al.}(2010)\citenamefont{Dreher, Donhauser,
  Daeubler, Glunk, Rapp, Schoch, Sauer, and Limmer}}]{PRB81_245202}
\bibinfo{author}{\bibfnamefont{L.}~\bibnamefont{Dreher}},
  \bibinfo{author}{\bibfnamefont{D.}~\bibnamefont{Donhauser}},
  \bibinfo{author}{\bibfnamefont{J.}~\bibnamefont{Daeubler}},
  \bibinfo{author}{\bibfnamefont{M.}~\bibnamefont{Glunk}},
  \bibinfo{author}{\bibfnamefont{C.}~\bibnamefont{Rapp}},
  \bibinfo{author}{\bibfnamefont{W.}~\bibnamefont{Schoch}},
  \bibinfo{author}{\bibfnamefont{R.}~\bibnamefont{Sauer}}, \bibnamefont{and}
  \bibinfo{author}{\bibfnamefont{W.}~\bibnamefont{Limmer}},
  \bibinfo{journal}{Phys. Rev. B} \textbf{\bibinfo{volume}{81}},
  \bibinfo{pages}{245202} (\bibinfo{year}{2010}).

\bibitem[{\citenamefont{Rushforth et~al.}(2007)\citenamefont{Rushforth,
  Vyborny, King, Edmonds, Campion, Foxon, Wunderlich, Irvine, Vasek, Novak
  et~al.}}]{PRL99_147207}
\bibinfo{author}{\bibfnamefont{A.~W.} \bibnamefont{Rushforth}},
  \bibinfo{author}{\bibfnamefont{K.}~\bibnamefont{Vyborny}},
  \bibinfo{author}{\bibfnamefont{C.~S.} \bibnamefont{King}},
  \bibinfo{author}{\bibfnamefont{K.~W.} \bibnamefont{Edmonds}},
  \bibinfo{author}{\bibfnamefont{R.~P.} \bibnamefont{Campion}},
  \bibinfo{author}{\bibfnamefont{C.~T.} \bibnamefont{Foxon}},
  \bibinfo{author}{\bibfnamefont{J.}~\bibnamefont{Wunderlich}},
  \bibinfo{author}{\bibfnamefont{A.~C.} \bibnamefont{Irvine}},
  \bibinfo{author}{\bibfnamefont{P.}~\bibnamefont{Vasek}},
  \bibinfo{author}{\bibfnamefont{V.}~\bibnamefont{Novak}},
  \bibinfo{author}{\bibfnamefont{K.}~\bibnamefont{Olejnik}},
  \bibinfo{author}{\bibfnamefont{J.}~\bibnamefont{Sinova}},
  \bibinfo{author}{\bibfnamefont{T.}~\bibnamefont{Jungwirth}},
  \bibnamefont{and} \bibinfo{author}{\bibfnamefont{B.~L.}
  \bibnamefont{Gallagher}}, \bibinfo{journal}{Phys. Rev. Lett.}
  \textbf{\bibinfo{volume}{99}}, \bibinfo{pages}{147207}
  (\bibinfo{year}{2007}).

\bibitem[{\citenamefont{Limmer et~al.}(2008)\citenamefont{Limmer, Daeubler,
  Dreher, Glunk, Schoch, Schwaiger, and Sauer}}]{PRB77_205210}
\bibinfo{author}{\bibfnamefont{W.}~\bibnamefont{Limmer}},
  \bibinfo{author}{\bibfnamefont{J.}~\bibnamefont{Daeubler}},
  \bibinfo{author}{\bibfnamefont{L.}~\bibnamefont{Dreher}},
  \bibinfo{author}{\bibfnamefont{M.}~\bibnamefont{Glunk}},
  \bibinfo{author}{\bibfnamefont{W.}~\bibnamefont{Schoch}},
  \bibinfo{author}{\bibfnamefont{S.}~\bibnamefont{Schwaiger}},
  \bibnamefont{and} \bibinfo{author}{\bibfnamefont{R.}~\bibnamefont{Sauer}},
  \bibinfo{journal}{Phys. Rev. B} \textbf{\bibinfo{volume}{77}},
  \bibinfo{pages}{205210} (\bibinfo{year}{2008}).

\bibitem[{\citenamefont{Pu et~al.}(2006)\citenamefont{Pu, Johnston-Halperin,
  Awschalom, and Shi}}]{PRL97_36601}
\bibinfo{author}{\bibfnamefont{Y.}~\bibnamefont{Pu}},
  \bibinfo{author}{\bibfnamefont{E.}~\bibnamefont{Johnston-Halperin}},
  \bibinfo{author}{\bibfnamefont{D.~D.} \bibnamefont{Awschalom}},
  \bibnamefont{and} \bibinfo{author}{\bibfnamefont{J.}~\bibnamefont{Shi}},
  \bibinfo{journal}{Phys. Rev. Lett.} \textbf{\bibinfo{volume}{97}},
  \bibinfo{pages}{036601} (\bibinfo{year}{2006}).

\bibitem[{\citenamefont{Jungwirth et~al.}(2006)\citenamefont{Jungwirth, Sinova,
  Masek, Kucera, and MacDonald}}]{RMP78_809}
\bibinfo{author}{\bibfnamefont{T.}~\bibnamefont{Jungwirth}},
  \bibinfo{author}{\bibfnamefont{J.}~\bibnamefont{Sinova}},
  \bibinfo{author}{\bibfnamefont{J.}~\bibnamefont{Masek}},
  \bibinfo{author}{\bibfnamefont{J.}~\bibnamefont{Kucera}}, \bibnamefont{and}
  \bibinfo{author}{\bibfnamefont{A.~H.} \bibnamefont{MacDonald}},
  \bibinfo{journal}{Rev. Mod. Phys.} \textbf{\bibinfo{volume}{78}},
  \bibinfo{pages}{809} (\bibinfo{year}{2006}).

\bibitem[{\citenamefont{Ohno}(2010)}]{NM9_952}
\bibinfo{author}{\bibfnamefont{H.}~\bibnamefont{Ohno}}, \bibinfo{journal}{Nat.
  Mater.} \textbf{\bibinfo{volume}{9}}, \bibinfo{pages}{952}
  (\bibinfo{year}{2010}).

\bibitem[{\citenamefont{Dietl}(2010)}]{NM9_965}
\bibinfo{author}{\bibfnamefont{T.}~\bibnamefont{Dietl}}, \bibinfo{journal}{Nat.
  Mater.} \textbf{\bibinfo{volume}{9}}, \bibinfo{pages}{965}
  (\bibinfo{year}{2010}).

\bibitem[{\citenamefont{Jaworski et~al.}(2010)\citenamefont{Jaworski, Yang,
  Mack, Awschalom, Heremans, and Myers}}]{NM9_898}
\bibinfo{author}{\bibfnamefont{C.~M.} \bibnamefont{Jaworski}},
  \bibinfo{author}{\bibfnamefont{J.}~\bibnamefont{Yang}},
  \bibinfo{author}{\bibfnamefont{S.}~\bibnamefont{Mack}},
  \bibinfo{author}{\bibfnamefont{D.~D.} \bibnamefont{Awschalom}},
  \bibinfo{author}{\bibfnamefont{J.~P.} \bibnamefont{Heremans}},
  \bibnamefont{and} \bibinfo{author}{\bibfnamefont{R.~C.} \bibnamefont{Myers}},
  \bibinfo{journal}{Nat. Mater.} \textbf{\bibinfo{volume}{9}},
  \bibinfo{pages}{898} (\bibinfo{year}{2010}).

\bibitem[{\citenamefont{Mark et~al.}(2011)\citenamefont{Mark, D{\"u}rrenfeld,
  Pappert, Ebel, Brunner, Gould, and Molenkamp}}]{PRL106_57204}
\bibinfo{author}{\bibfnamefont{S.}~\bibnamefont{Mark}},
  \bibinfo{author}{\bibfnamefont{P.}~\bibnamefont{D{\"u}rrenfeld}},
  \bibinfo{author}{\bibfnamefont{K.}~\bibnamefont{Pappert}},
  \bibinfo{author}{\bibfnamefont{L.}~\bibnamefont{Ebel}},
  \bibinfo{author}{\bibfnamefont{K.}~\bibnamefont{Brunner}},
  \bibinfo{author}{\bibfnamefont{C.}~\bibnamefont{Gould}}, \bibnamefont{and}
  \bibinfo{author}{\bibfnamefont{L.~W.} \bibnamefont{Molenkamp}},
  \bibinfo{journal}{Phys. Rev. Lett.} \textbf{\bibinfo{volume}{106}},
  \bibinfo{pages}{057204} (\bibinfo{year}{2011}).

\bibitem[{\citenamefont{Dietl et~al.}(2001)\citenamefont{Dietl, Ohno, and
  Matsukura}}]{PRB63_195205}
\bibinfo{author}{\bibfnamefont{T.}~\bibnamefont{Dietl}},
  \bibinfo{author}{\bibfnamefont{H.}~\bibnamefont{Ohno}}, \bibnamefont{and}
  \bibinfo{author}{\bibfnamefont{F.}~\bibnamefont{Matsukura}},
  \bibinfo{journal}{Phys. Rev. B} \textbf{\bibinfo{volume}{63}},
  \bibinfo{pages}{195205} (\bibinfo{year}{2001}).

\bibitem[{\citenamefont{Glunk et~al.}(2009)\citenamefont{Glunk, Daeubler,
  Dreher, Schwaiger, Schoch, Sauer, Limmer, Brandlmaier, Goennenwein, Bihler
  et~al.}}]{PRB79_195206}
\bibinfo{author}{\bibfnamefont{M.}~\bibnamefont{Glunk}},
  \bibinfo{author}{\bibfnamefont{J.}~\bibnamefont{Daeubler}},
  \bibinfo{author}{\bibfnamefont{L.}~\bibnamefont{Dreher}},
  \bibinfo{author}{\bibfnamefont{S.}~\bibnamefont{Schwaiger}},
  \bibinfo{author}{\bibfnamefont{W.}~\bibnamefont{Schoch}},
  \bibinfo{author}{\bibfnamefont{R.}~\bibnamefont{Sauer}},
  \bibinfo{author}{\bibfnamefont{W.}~\bibnamefont{Limmer}},
  \bibinfo{author}{\bibfnamefont{A.}~\bibnamefont{Brandlmaier}},
  \bibinfo{author}{\bibfnamefont{S.~T.~B.} \bibnamefont{Goennenwein}},
  \bibinfo{author}{\bibfnamefont{C.}~\bibnamefont{Bihler}}, \bibnamefont{and}
  \bibinfo{author}{\bibfnamefont{M.~S.} \bibnamefont{Brandt}},
  \bibinfo{journal}{Phys. Rev. B} \textbf{\bibinfo{volume}{79}},
  \bibinfo{pages}{195206} (\bibinfo{year}{2009}).

\bibitem[{\citenamefont{Goennenwein et~al.}(2008)\citenamefont{Goennenwein,
  Althammer, Bihler, Brandlmaier, Gepraegs, Opel, Schoch, Limmer, Gross, and
  Brandt}}]{PSS2_96}
\bibinfo{author}{\bibfnamefont{S.~T.~B.} \bibnamefont{Goennenwein}},
  \bibinfo{author}{\bibfnamefont{M.}~\bibnamefont{Althammer}},
  \bibinfo{author}{\bibfnamefont{C.}~\bibnamefont{Bihler}},
  \bibinfo{author}{\bibfnamefont{A.}~\bibnamefont{Brandlmaier}},
  \bibinfo{author}{\bibfnamefont{S.}~\bibnamefont{Gepraegs}},
  \bibinfo{author}{\bibfnamefont{M.}~\bibnamefont{Opel}},
  \bibinfo{author}{\bibfnamefont{W.}~\bibnamefont{Schoch}},
  \bibinfo{author}{\bibfnamefont{W.}~\bibnamefont{Limmer}},
  \bibinfo{author}{\bibfnamefont{R.}~\bibnamefont{Gross}}, \bibnamefont{and}
  \bibinfo{author}{\bibfnamefont{M.~S.} \bibnamefont{Brandt}},
  \bibinfo{journal}{Phys. Status Solidi {(RRL)}} \textbf{\bibinfo{volume}{2}},
  \bibinfo{pages}{96} (\bibinfo{year}{2008}).

\bibitem[{\citenamefont{DeRanieri et~al.}(2008)\citenamefont{DeRanieri,
  Rushforth, Vyborny, Rana, Ahmad, Campion, Foxon, Gallagher, Irvine,
  Wunderlich et~al.}}]{NJoP10_65003}
\bibinfo{author}{\bibfnamefont{E.}~\bibnamefont{DeRanieri}},
  \bibinfo{author}{\bibfnamefont{A.~W.} \bibnamefont{Rushforth}},
  \bibinfo{author}{\bibfnamefont{K.}~\bibnamefont{Vyborny}},
  \bibinfo{author}{\bibfnamefont{U.}~\bibnamefont{Rana}},
  \bibinfo{author}{\bibfnamefont{E.}~\bibnamefont{Ahmad}},
  \bibinfo{author}{\bibfnamefont{R.~P.} \bibnamefont{Campion}},
  \bibinfo{author}{\bibfnamefont{C.~T.} \bibnamefont{Foxon}},
  \bibinfo{author}{\bibfnamefont{B.~L.} \bibnamefont{Gallagher}},
  \bibinfo{author}{\bibfnamefont{A.~C.} \bibnamefont{Irvine}},
  \bibinfo{author}{\bibfnamefont{J.}~\bibnamefont{Wunderlich}},
  \bibnamefont{and}
  \bibinfo{author}{\bibfnamefont{T.}~\bibnamefont{Jungwirth}},
  \bibinfo{journal}{New J. Phys.} \textbf{\bibinfo{volume}{10}},
  \bibinfo{pages}{065003} (\bibinfo{year}{2008}).

\bibitem[{\citenamefont{Bihler et~al.}(2008)\citenamefont{Bihler, Althammer,
  Brandlmaier, Gepr\"{a}gs, Weiler, Opel, Schoch, Limmer, Gross, Brandt
  et~al.}}]{PRB78_45203}
\bibinfo{author}{\bibfnamefont{C.}~\bibnamefont{Bihler}},
  \bibinfo{author}{\bibfnamefont{M.}~\bibnamefont{Althammer}},
  \bibinfo{author}{\bibfnamefont{A.}~\bibnamefont{Brandlmaier}},
  \bibinfo{author}{\bibfnamefont{S.}~\bibnamefont{Gepr\"{a}gs}},
  \bibinfo{author}{\bibfnamefont{M.}~\bibnamefont{Weiler}},
  \bibinfo{author}{\bibfnamefont{M.}~\bibnamefont{Opel}},
  \bibinfo{author}{\bibfnamefont{W.}~\bibnamefont{Schoch}},
  \bibinfo{author}{\bibfnamefont{W.}~\bibnamefont{Limmer}},
  \bibinfo{author}{\bibfnamefont{R.}~\bibnamefont{Gross}},
  \bibinfo{author}{\bibfnamefont{M.~S.} \bibnamefont{Brandt}},
  \bibnamefont{and} \bibinfo{author}{\bibfnamefont{S.~T.~B.}
  \bibnamefont{Goennenwein}}, \bibinfo{journal}{Phys. Rev. B}
  \textbf{\bibinfo{volume}{78}}, \bibinfo{pages}{045203}
  (\bibinfo{year}{2008}).

\bibitem[{\citenamefont{Rado}(1961)}]{JAP32_129}
\bibinfo{author}{\bibfnamefont{G.~T.} \bibnamefont{Rado}}, \bibinfo{journal}{J.
  Appl. Phys.} \textbf{\bibinfo{volume}{32}}, \bibinfo{pages}{S129}
  (\bibinfo{year}{1961}).

\bibitem[{\citenamefont{Rubinstein et~al.}(2002)\citenamefont{Rubinstein,
  Hanbicki, Lubitz, Osofsky, Krebs, and Jonker}}]{JoMaMM250_164}
\bibinfo{author}{\bibfnamefont{M.}~\bibnamefont{Rubinstein}},
  \bibinfo{author}{\bibfnamefont{A.}~\bibnamefont{Hanbicki}},
  \bibinfo{author}{\bibfnamefont{P.}~\bibnamefont{Lubitz}},
  \bibinfo{author}{\bibfnamefont{M.}~\bibnamefont{Osofsky}},
  \bibinfo{author}{\bibfnamefont{J.}~\bibnamefont{Krebs}}, \bibnamefont{and}
  \bibinfo{author}{\bibfnamefont{B.}~\bibnamefont{Jonker}},
  \bibinfo{journal}{J. Magn. Magn. Mater.} \textbf{\bibinfo{volume}{250}},
  \bibinfo{pages}{164} (\bibinfo{year}{2002}).

\bibitem[{\citenamefont{Sasaki et~al.}(2002)\citenamefont{Sasaki, Liu, Furdyna,
  Palczewska, Szczytko, and Twardowski}}]{JAP91_7484}
\bibinfo{author}{\bibfnamefont{Y.}~\bibnamefont{Sasaki}},
  \bibinfo{author}{\bibfnamefont{X.}~\bibnamefont{Liu}},
  \bibinfo{author}{\bibfnamefont{J.~K.} \bibnamefont{Furdyna}},
  \bibinfo{author}{\bibfnamefont{M.}~\bibnamefont{Palczewska}},
  \bibinfo{author}{\bibfnamefont{J.}~\bibnamefont{Szczytko}}, \bibnamefont{and}
  \bibinfo{author}{\bibfnamefont{A.}~\bibnamefont{Twardowski}},
  \bibinfo{journal}{J. Appl. Phys.} \textbf{\bibinfo{volume}{91}},
  \bibinfo{pages}{7484} (\bibinfo{year}{2002}).

\bibitem[{\citenamefont{Liu et~al.}(2003)\citenamefont{Liu, Sasaki, and
  Furdyna}}]{PRB67_205204}
\bibinfo{author}{\bibfnamefont{X.}~\bibnamefont{Liu}},
  \bibinfo{author}{\bibfnamefont{Y.}~\bibnamefont{Sasaki}}, \bibnamefont{and}
  \bibinfo{author}{\bibfnamefont{J.~K.} \bibnamefont{Furdyna}},
  \bibinfo{journal}{Phys. Rev. B} \textbf{\bibinfo{volume}{67}},
  \bibinfo{pages}{205204} (\bibinfo{year}{2003}).

\bibitem[{\citenamefont{Bihler et~al.}(2006)\citenamefont{Bihler, Huebl,
  Brandt, Goennenwein, Reinwald, Wurstbauer, Doppe, Weiss, and
  Wegscheider}}]{APL89_012507}
\bibinfo{author}{\bibfnamefont{C.}~\bibnamefont{Bihler}},
  \bibinfo{author}{\bibfnamefont{H.}~\bibnamefont{Huebl}},
  \bibinfo{author}{\bibfnamefont{M.~S.} \bibnamefont{Brandt}},
  \bibinfo{author}{\bibfnamefont{S.~T.~B.} \bibnamefont{Goennenwein}},
  \bibinfo{author}{\bibfnamefont{M.}~\bibnamefont{Reinwald}},
  \bibinfo{author}{\bibfnamefont{U.}~\bibnamefont{Wurstbauer}},
  \bibinfo{author}{\bibfnamefont{M.}~\bibnamefont{Doppe}},
  \bibinfo{author}{\bibfnamefont{D.}~\bibnamefont{Weiss}}, \bibnamefont{and}
  \bibinfo{author}{\bibfnamefont{W.}~\bibnamefont{Wegscheider}},
  \bibinfo{journal}{Appl. Phys. Lett.} \textbf{\bibinfo{volume}{89}},
  \bibinfo{pages}{012507} (\bibinfo{year}{2006}).

\bibitem[{\citenamefont{Liu and Furdyna}(2006)}]{LiuFurdyna_FMR}
\bibinfo{author}{\bibfnamefont{X.}~\bibnamefont{Liu}} \bibnamefont{and}
  \bibinfo{author}{\bibfnamefont{J.~K.} \bibnamefont{Furdyna}},
  \bibinfo{journal}{J. Phys.: Cond. Mat.} \textbf{\bibinfo{volume}{18}},
  \bibinfo{pages}{R245} (\bibinfo{year}{2006}).

\bibitem[{\citenamefont{Khazen et~al.}(2008{\natexlab{a}})\citenamefont{Khazen,
  von Bardeleben, Cantin, Thevenard, Largeau, Mauguin, and
  Lema{\^i}tre}}]{PRB77_165204}
\bibinfo{author}{\bibfnamefont{K.}~\bibnamefont{Khazen}},
  \bibinfo{author}{\bibfnamefont{H.~J.} \bibnamefont{von Bardeleben}},
  \bibinfo{author}{\bibfnamefont{J.~L.} \bibnamefont{Cantin}},
  \bibinfo{author}{\bibfnamefont{L.}~\bibnamefont{Thevenard}},
  \bibinfo{author}{\bibfnamefont{L.}~\bibnamefont{Largeau}},
  \bibinfo{author}{\bibfnamefont{O.}~\bibnamefont{Mauguin}}, \bibnamefont{and}
  \bibinfo{author}{\bibfnamefont{A.}~\bibnamefont{Lema{\^i}tre}},
  \bibinfo{journal}{Phys. Rev. B} \textbf{\bibinfo{volume}{77}},
  \bibinfo{pages}{165204} (\bibinfo{year}{2008}{\natexlab{a}}).

\bibitem[{\citenamefont{Sandweg et~al.}(2010)\citenamefont{Sandweg, Kajiwara,
  Ando, Saitoh, and Hillebrands}}]{APL97_252504}
\bibinfo{author}{\bibfnamefont{C.~W.} \bibnamefont{Sandweg}},
  \bibinfo{author}{\bibfnamefont{Y.}~\bibnamefont{Kajiwara}},
  \bibinfo{author}{\bibfnamefont{K.}~\bibnamefont{Ando}},
  \bibinfo{author}{\bibfnamefont{E.}~\bibnamefont{Saitoh}}, \bibnamefont{and}
  \bibinfo{author}{\bibfnamefont{B.}~\bibnamefont{Hillebrands}},
  \bibinfo{journal}{Appl. Phys. Lett.} \textbf{\bibinfo{volume}{97}},
  \bibinfo{pages}{252504} (\bibinfo{year}{2010}).

\bibitem[{\citenamefont{Sandweg et~al.}(2011)\citenamefont{Sandweg, Kajiwara,
  Chumak, Serga, Vasyuchka, Jungfleisch, Saitoh, and
  Hillebrands}}]{PRL106_216601}
\bibinfo{author}{\bibfnamefont{C.~W.} \bibnamefont{Sandweg}},
  \bibinfo{author}{\bibfnamefont{Y.}~\bibnamefont{Kajiwara}},
  \bibinfo{author}{\bibfnamefont{A.~V.} \bibnamefont{Chumak}},
  \bibinfo{author}{\bibfnamefont{A.~A.} \bibnamefont{Serga}},
  \bibinfo{author}{\bibfnamefont{V.~I.} \bibnamefont{Vasyuchka}},
  \bibinfo{author}{\bibfnamefont{M.~B.} \bibnamefont{Jungfleisch}},
  \bibinfo{author}{\bibfnamefont{E.}~\bibnamefont{Saitoh}}, \bibnamefont{and}
  \bibinfo{author}{\bibfnamefont{B.}~\bibnamefont{Hillebrands}},
  \bibinfo{journal}{Phys. Rev. Lett.} \textbf{\bibinfo{volume}{106}},
  \bibinfo{pages}{216601} (\bibinfo{year}{2011}).

\bibitem[{\citenamefont{Bauer and Tserkovnyak}(2011)}]{P4_40}
\bibinfo{author}{\bibfnamefont{G.~E.~W.} \bibnamefont{Bauer}} \bibnamefont{and}
  \bibinfo{author}{\bibfnamefont{Y.}~\bibnamefont{Tserkovnyak}},
  \bibinfo{journal}{Physics} \textbf{\bibinfo{volume}{4}}, \bibinfo{pages}{40}
  (\bibinfo{year}{2011}).

\bibitem[{\citenamefont{Czeschka et~al.}(2011)\citenamefont{Czeschka, Dreher,
  Brandt, Weiler, Althammer, Imort, Reiss, Thomas, Schoch, Limmer
  et~al.}}]{PRL107_46601}
\bibinfo{author}{\bibfnamefont{F.~D.} \bibnamefont{Czeschka}},
  \bibinfo{author}{\bibfnamefont{L.}~\bibnamefont{Dreher}},
  \bibinfo{author}{\bibfnamefont{M.~S.} \bibnamefont{Brandt}},
  \bibinfo{author}{\bibfnamefont{M.}~\bibnamefont{Weiler}},
  \bibinfo{author}{\bibfnamefont{M.}~\bibnamefont{Althammer}},
  \bibinfo{author}{\bibfnamefont{I.~M.} \bibnamefont{Imort}},
  \bibinfo{author}{\bibfnamefont{G.}~\bibnamefont{Reiss}},
  \bibinfo{author}{\bibfnamefont{A.}~\bibnamefont{Thomas}},
  \bibinfo{author}{\bibfnamefont{W.}~\bibnamefont{Schoch}},
  \bibinfo{author}{\bibfnamefont{W.}~\bibnamefont{Limmer}},
  \bibinfo{author}{\bibfnamefont{H.}~\bibnamefont{Huebl}},
  \bibinfo{author}{\bibfnamefont{R.}~\bibnamefont{Gross}}, \bibnamefont{and}
  \bibinfo{author}{\bibfnamefont{S.~T.~B.} \bibnamefont{Goennenwein}},
  \bibinfo{journal}{Phys. Rev. Lett.} \textbf{\bibinfo{volume}{107}},
  \bibinfo{pages}{046601} (\bibinfo{year}{2011}).

\bibitem[{\citenamefont{Tserkovnyak et~al.}(2002)\citenamefont{Tserkovnyak,
  Brataas, and Bauer}}]{PRL88_117601}
\bibinfo{author}{\bibfnamefont{Y.}~\bibnamefont{Tserkovnyak}},
  \bibinfo{author}{\bibfnamefont{A.}~\bibnamefont{Brataas}}, \bibnamefont{and}
  \bibinfo{author}{\bibfnamefont{G.~E.~W.} \bibnamefont{Bauer}},
  \bibinfo{journal}{Phys. Rev. Lett.} \textbf{\bibinfo{volume}{88}},
  \bibinfo{pages}{117601} (\bibinfo{year}{2002}).

\bibitem[{\citenamefont{Mosendz et~al.}(2010)\citenamefont{Mosendz, Pearson,
  Fradin, Bauer, Bader, and Hoffmann}}]{PRL104_46601}
\bibinfo{author}{\bibfnamefont{O.}~\bibnamefont{Mosendz}},
  \bibinfo{author}{\bibfnamefont{J.~E.} \bibnamefont{Pearson}},
  \bibinfo{author}{\bibfnamefont{F.~Y.} \bibnamefont{Fradin}},
  \bibinfo{author}{\bibfnamefont{G.~E.~W.} \bibnamefont{Bauer}},
  \bibinfo{author}{\bibfnamefont{S.~D.} \bibnamefont{Bader}}, \bibnamefont{and}
  \bibinfo{author}{\bibfnamefont{A.}~\bibnamefont{Hoffmann}},
  \bibinfo{journal}{Phys. Rev. Lett.} \textbf{\bibinfo{volume}{104}},
  \bibinfo{pages}{046601} (\bibinfo{year}{2010}).

\bibitem[{\citenamefont{Tserkovnyak et~al.}(2005)\citenamefont{Tserkovnyak,
  Brataas, Bauer, and Halperin}}]{RMP77_1375}
\bibinfo{author}{\bibfnamefont{Y.}~\bibnamefont{Tserkovnyak}},
  \bibinfo{author}{\bibfnamefont{A.}~\bibnamefont{Brataas}},
  \bibinfo{author}{\bibfnamefont{G.~E.~W.} \bibnamefont{Bauer}},
  \bibnamefont{and} \bibinfo{author}{\bibfnamefont{B.~I.}
  \bibnamefont{Halperin}}, \bibinfo{journal}{Rev. Mod. Phys.}
  \textbf{\bibinfo{volume}{77}}, \bibinfo{pages}{1375} (\bibinfo{year}{2005}).

\bibitem[{\citenamefont{Goennenwein et~al.}(2003)\citenamefont{Goennenwein,
  Graf, Wassner, Brandt, Stutzmann, Philipp, Gross, Krieger, Zurn, Ziemann
  et~al.}}]{APL82_730}
\bibinfo{author}{\bibfnamefont{S.~T.~B.} \bibnamefont{Goennenwein}},
  \bibinfo{author}{\bibfnamefont{T.}~\bibnamefont{Graf}},
  \bibinfo{author}{\bibfnamefont{T.}~\bibnamefont{Wassner}},
  \bibinfo{author}{\bibfnamefont{M.~S.} \bibnamefont{Brandt}},
  \bibinfo{author}{\bibfnamefont{M.}~\bibnamefont{Stutzmann}},
  \bibinfo{author}{\bibfnamefont{J.~B.} \bibnamefont{Philipp}},
  \bibinfo{author}{\bibfnamefont{R.}~\bibnamefont{Gross}},
  \bibinfo{author}{\bibfnamefont{M.}~\bibnamefont{Krieger}},
  \bibinfo{author}{\bibfnamefont{K.}~\bibnamefont{Zurn}},
  \bibinfo{author}{\bibfnamefont{P.}~\bibnamefont{Ziemann}},
  \bibinfo{author}{\bibfnamefont{A.}~\bibnamefont{Koeder}},
  \bibinfo{author}{\bibfnamefont{S.}~\bibnamefont{Frank}},
  \bibinfo{author}{\bibfnamefont{W.}~\bibnamefont{Schoch}}, \bibnamefont{and}
  \bibinfo{author}{\bibfnamefont{A.}~\bibnamefont{Waag}},
  \bibinfo{journal}{Appl. Phys. Lett.} \textbf{\bibinfo{volume}{82}},
  \bibinfo{pages}{730} (\bibinfo{year}{2003}).

\bibitem[{\citenamefont{Sasaki et~al.}(2003)\citenamefont{Sasaki, Liu,
  Wojtowicz, and Furdyna}}]{JoS16_143}
\bibinfo{author}{\bibfnamefont{Y.}~\bibnamefont{Sasaki}},
  \bibinfo{author}{\bibfnamefont{X.}~\bibnamefont{Liu}},
  \bibinfo{author}{\bibfnamefont{T.}~\bibnamefont{Wojtowicz}},
  \bibnamefont{and} \bibinfo{author}{\bibfnamefont{J.~K.}
  \bibnamefont{Furdyna}}, \bibinfo{journal}{J. Supercond. Novel Magnetism}
  \textbf{\bibinfo{volume}{16}}, \bibinfo{pages}{143} (\bibinfo{year}{2003}).

\bibitem[{\citenamefont{Rappoport et~al.}(2004)\citenamefont{Rappoport,
  Redli\ifmmode~\acute{n}\else \'{n}\fi{}ski, Liu, Zar\'and, Furdyna, and
  Jank\'o}}]{PRB69_125213}
\bibinfo{author}{\bibfnamefont{T.~G.} \bibnamefont{Rappoport}},
  \bibinfo{author}{\bibfnamefont{P.}~\bibnamefont{Redli\ifmmode~\acute{n}\else
  \'{n}\fi{}ski}}, \bibinfo{author}{\bibfnamefont{X.}~\bibnamefont{Liu}},
  \bibinfo{author}{\bibfnamefont{G.}~\bibnamefont{Zar\'and}},
  \bibinfo{author}{\bibfnamefont{J.~K.} \bibnamefont{Furdyna}},
  \bibnamefont{and} \bibinfo{author}{\bibfnamefont{B.}~\bibnamefont{Jank\'o}},
  \bibinfo{journal}{Phys. Rev. B} \textbf{\bibinfo{volume}{69}},
  \bibinfo{pages}{125213} (\bibinfo{year}{2004}).

\bibitem[{\citenamefont{Zhou et~al.}(2007)\citenamefont{Zhou, Cho, Ge, Liu,
  Dobrowolska, and Furdyna}}]{MITo43_3019}
\bibinfo{author}{\bibfnamefont{Y.-Y.} \bibnamefont{Zhou}},
  \bibinfo{author}{\bibfnamefont{Y.-J.} \bibnamefont{Cho}},
  \bibinfo{author}{\bibfnamefont{Z.}~\bibnamefont{Ge}},
  \bibinfo{author}{\bibfnamefont{X.}~\bibnamefont{Liu}},
  \bibinfo{author}{\bibfnamefont{M.}~\bibnamefont{Dobrowolska}},
  \bibnamefont{and} \bibinfo{author}{\bibfnamefont{J.}~\bibnamefont{Furdyna}},
  \bibinfo{journal}{IEEE Trans. Magn.} \textbf{\bibinfo{volume}{43}},
  \bibinfo{pages}{3019} (\bibinfo{year}{2007}).

\bibitem[{\citenamefont{Liu et~al.}(2007)\citenamefont{Liu, Zhou, and
  Furdyna}}]{PRB75_195220}
\bibinfo{author}{\bibfnamefont{X.}~\bibnamefont{Liu}},
  \bibinfo{author}{\bibfnamefont{Y.~Y.} \bibnamefont{Zhou}}, \bibnamefont{and}
  \bibinfo{author}{\bibfnamefont{J.~K.} \bibnamefont{Furdyna}},
  \bibinfo{journal}{Phys. Rev. B} \textbf{\bibinfo{volume}{75}},
  \bibinfo{pages}{195220} (\bibinfo{year}{2007}).

\bibitem[{\citenamefont{Bihler et~al.}(2009)\citenamefont{Bihler, Schoch,
  Limmer, Goennenwein, and Brandt}}]{PRB79_45205}
\bibinfo{author}{\bibfnamefont{C.}~\bibnamefont{Bihler}},
  \bibinfo{author}{\bibfnamefont{W.}~\bibnamefont{Schoch}},
  \bibinfo{author}{\bibfnamefont{W.}~\bibnamefont{Limmer}},
  \bibinfo{author}{\bibfnamefont{S.~T.~B.} \bibnamefont{Goennenwein}},
  \bibnamefont{and} \bibinfo{author}{\bibfnamefont{M.~S.}
  \bibnamefont{Brandt}}, \bibinfo{journal}{Phys. Rev. B}
  \textbf{\bibinfo{volume}{79}}, \bibinfo{pages}{045205}
  (\bibinfo{year}{2009}).

\bibitem[{\citenamefont{Hoekstra et~al.}(1977)\citenamefont{Hoekstra, van
  Stapele, and Robertson}}]{JAP48_382}
\bibinfo{author}{\bibfnamefont{B.}~\bibnamefont{Hoekstra}},
  \bibinfo{author}{\bibfnamefont{R.~P.} \bibnamefont{van Stapele}},
  \bibnamefont{and} \bibinfo{author}{\bibfnamefont{J.~M.}
  \bibnamefont{Robertson}}, \bibinfo{journal}{J. Appl. Phys.}
  \textbf{\bibinfo{volume}{48}}, \bibinfo{pages}{382} (\bibinfo{year}{1977}).

\bibitem[{\citenamefont{Hoekstra}(1978)}]{Hoekstra_diss}
\bibinfo{author}{\bibfnamefont{B.}~\bibnamefont{Hoekstra}}, Ph.D. thesis,
  \bibinfo{school}{Philips Research Laboratories}, \bibinfo{address}{Eindhoven}
  (\bibinfo{year}{1978}).

\bibitem[{\citenamefont{Limmer et~al.}(2005)\citenamefont{Limmer, Koeder,
  Frank, Avrutin, Schoch, Sauer, Zuern, Eisenmenger, Ziemann, Peiner
  et~al.}}]{PRB71_205213}
\bibinfo{author}{\bibfnamefont{W.}~\bibnamefont{Limmer}},
  \bibinfo{author}{\bibfnamefont{A.}~\bibnamefont{Koeder}},
  \bibinfo{author}{\bibfnamefont{S.}~\bibnamefont{Frank}},
  \bibinfo{author}{\bibfnamefont{V.}~\bibnamefont{Avrutin}},
  \bibinfo{author}{\bibfnamefont{W.}~\bibnamefont{Schoch}},
  \bibinfo{author}{\bibfnamefont{R.}~\bibnamefont{Sauer}},
  \bibinfo{author}{\bibfnamefont{K.}~\bibnamefont{Zuern}},
  \bibinfo{author}{\bibfnamefont{J.}~\bibnamefont{Eisenmenger}},
  \bibinfo{author}{\bibfnamefont{P.}~\bibnamefont{Ziemann}},
  \bibinfo{author}{\bibfnamefont{E.}~\bibnamefont{Peiner}}, \bibnamefont{and}
  \bibinfo{author}{\bibfnamefont{A.}~\bibnamefont{Waag}},
  \bibinfo{journal}{Phys. Rev. B} \textbf{\bibinfo{volume}{71}},
  \bibinfo{pages}{205213} (\bibinfo{year}{2005}).

\bibitem[{\citenamefont{Farle}(1998)}]{RoPiP61_755}
\bibinfo{author}{\bibfnamefont{M.}~\bibnamefont{Farle}}, \bibinfo{journal}{Rep.
  Prog. Phys.} \textbf{\bibinfo{volume}{61}}, \bibinfo{pages}{755}
  (\bibinfo{year}{1998}).

\bibitem[{\citenamefont{Limmer et~al.}(2006)\citenamefont{Limmer, Glunk,
  Daeubler, Hummel, Schoch, Sauer, Bihler, Huebl, Brandt, and
  Goennenwein}}]{PRB74_205205}
\bibinfo{author}{\bibfnamefont{W.}~\bibnamefont{Limmer}},
  \bibinfo{author}{\bibfnamefont{M.}~\bibnamefont{Glunk}},
  \bibinfo{author}{\bibfnamefont{J.}~\bibnamefont{Daeubler}},
  \bibinfo{author}{\bibfnamefont{T.}~\bibnamefont{Hummel}},
  \bibinfo{author}{\bibfnamefont{W.}~\bibnamefont{Schoch}},
  \bibinfo{author}{\bibfnamefont{R.}~\bibnamefont{Sauer}},
  \bibinfo{author}{\bibfnamefont{C.}~\bibnamefont{Bihler}},
  \bibinfo{author}{\bibfnamefont{H.}~\bibnamefont{Huebl}},
  \bibinfo{author}{\bibfnamefont{M.~S.} \bibnamefont{Brandt}},
  \bibnamefont{and} \bibinfo{author}{\bibfnamefont{S.~T.~B.}
  \bibnamefont{Goennenwein}}, \bibinfo{journal}{Phys. Rev. B}
  \textbf{\bibinfo{volume}{74}}, \bibinfo{pages}{205205}
  (\bibinfo{year}{2006}).

\bibitem[{\citenamefont{Landau and Lifshitz}(1935)}]{LandauLifshitz}
\bibinfo{author}{\bibfnamefont{L.}~\bibnamefont{Landau}} \bibnamefont{and}
  \bibinfo{author}{\bibfnamefont{E.}~\bibnamefont{Lifshitz}},
  \bibinfo{journal}{Phys. Z. Sowjetunion} \textbf{\bibinfo{volume}{8}},
  \bibinfo{pages}{153} (\bibinfo{year}{1935}).

\bibitem[{\citenamefont{Gilbert}(2004)}]{gilbert_phenomenological_2004}
\bibinfo{author}{\bibfnamefont{T.}~\bibnamefont{Gilbert}},
  \bibinfo{journal}{{IEEE} Trans. Magn.} \textbf{\bibinfo{volume}{40}},
  \bibinfo{pages}{3443} (\bibinfo{year}{2004}).

\bibitem[{\citenamefont{Baselgia et~al.}(1988)\citenamefont{Baselgia, Warden,
  Waldner, Hutton, Drumheller, He, Wigen, and Marysko}}]{PRB38_2237}
\bibinfo{author}{\bibfnamefont{L.}~\bibnamefont{Baselgia}},
  \bibinfo{author}{\bibfnamefont{M.}~\bibnamefont{Warden}},
  \bibinfo{author}{\bibfnamefont{F.}~\bibnamefont{Waldner}},
  \bibinfo{author}{\bibfnamefont{S.~L.} \bibnamefont{Hutton}},
  \bibinfo{author}{\bibfnamefont{J.~E.} \bibnamefont{Drumheller}},
  \bibinfo{author}{\bibfnamefont{Y.~Q.} \bibnamefont{He}},
  \bibinfo{author}{\bibfnamefont{P.~E.} \bibnamefont{Wigen}}, \bibnamefont{and}
  \bibinfo{author}{\bibfnamefont{M.}~\bibnamefont{Marysko}},
  \bibinfo{journal}{Phys. Rev. B} \textbf{\bibinfo{volume}{38}},
  \bibinfo{pages}{2237} (\bibinfo{year}{1988}).

\bibitem[{\citenamefont{Smit and Beljiers}(1955)}]{PRR10_113}
\bibinfo{author}{\bibfnamefont{J.}~\bibnamefont{Smit}} \bibnamefont{and}
  \bibinfo{author}{\bibfnamefont{H.~G.} \bibnamefont{Beljiers}},
  \bibinfo{journal}{Philips Res. Rep.} \textbf{\bibinfo{volume}{10}},
  \bibinfo{pages}{113} (\bibinfo{year}{1955}).

\bibitem[{\citenamefont{Stone et~al.}(2010)\citenamefont{Stone, Dreher, Beeman,
  Yu, Brandt, and Dubon}}]{PRB81_205210}
\bibinfo{author}{\bibfnamefont{P.~R.} \bibnamefont{Stone}},
  \bibinfo{author}{\bibfnamefont{L.}~\bibnamefont{Dreher}},
  \bibinfo{author}{\bibfnamefont{J.~W.} \bibnamefont{Beeman}},
  \bibinfo{author}{\bibfnamefont{K.~M.} \bibnamefont{Yu}},
  \bibinfo{author}{\bibfnamefont{M.~S.} \bibnamefont{Brandt}},
  \bibnamefont{and} \bibinfo{author}{\bibfnamefont{O.~D.} \bibnamefont{Dubon}},
  \bibinfo{journal}{Phys. Rev. B} \textbf{\bibinfo{volume}{81}},
  \bibinfo{pages}{205210} (\bibinfo{year}{2010}).

\bibitem[{\citenamefont{Puszkarski}(1979)}]{PiSS9_191}
\bibinfo{author}{\bibfnamefont{H.}~\bibnamefont{Puszkarski}},
  \bibinfo{journal}{Prog. Surf. Sci.} \textbf{\bibinfo{volume}{9}},
  \bibinfo{pages}{191} (\bibinfo{year}{1979}).

\bibitem[{\citenamefont{Weiler et~al.}(2011)\citenamefont{Weiler, Dreher, Heeg,
  Huebl, Gross, Brandt, and Goennenwein}}]{PRL106_117601}
\bibinfo{author}{\bibfnamefont{M.}~\bibnamefont{Weiler}},
  \bibinfo{author}{\bibfnamefont{L.}~\bibnamefont{Dreher}},
  \bibinfo{author}{\bibfnamefont{C.}~\bibnamefont{Heeg}},
  \bibinfo{author}{\bibfnamefont{H.}~\bibnamefont{Huebl}},
  \bibinfo{author}{\bibfnamefont{R.}~\bibnamefont{Gross}},
  \bibinfo{author}{\bibfnamefont{M.~S.} \bibnamefont{Brandt}},
  \bibnamefont{and} \bibinfo{author}{\bibfnamefont{S.~T.~B.}
  \bibnamefont{Goennenwein}}, \bibinfo{journal}{Phys. Rev. Lett.}
  \textbf{\bibinfo{volume}{106}}, \bibinfo{pages}{117601}
  (\bibinfo{year}{2011}).

\bibitem[{\citenamefont{Dreher et~al.}(2012)\citenamefont{Dreher, Weiler,
  Pernpeintner, Huebl, Gross, Brandt, and Goennenwein}}]{PRB86_134415}
\bibinfo{author}{\bibfnamefont{L.}~\bibnamefont{Dreher}},
  \bibinfo{author}{\bibfnamefont{M.}~\bibnamefont{Weiler}},
  \bibinfo{author}{\bibfnamefont{M.}~\bibnamefont{Pernpeintner}},
  \bibinfo{author}{\bibfnamefont{H.}~\bibnamefont{Huebl}},
  \bibinfo{author}{\bibfnamefont{R.}~\bibnamefont{Gross}},
  \bibinfo{author}{\bibfnamefont{M.~S.} \bibnamefont{Brandt}},
  \bibnamefont{and} \bibinfo{author}{\bibfnamefont{S.~T.~B.}
  \bibnamefont{Goennenwein}}, \bibinfo{journal}{Phys. Rev. B}
  \textbf{\bibinfo{volume}{86}}, \bibinfo{pages}{134415}
  (\bibinfo{year}{2012}).

\bibitem[{\citenamefont{Wang et~al.}(2007)\citenamefont{Wang, Ren, Liu,
  Furdyna, Grimsditch, and Merlin}}]{PRB75_233308}
\bibinfo{author}{\bibfnamefont{D.~M.} \bibnamefont{Wang}},
  \bibinfo{author}{\bibfnamefont{Y.~H.} \bibnamefont{Ren}},
  \bibinfo{author}{\bibfnamefont{X.}~\bibnamefont{Liu}},
  \bibinfo{author}{\bibfnamefont{J.~K.} \bibnamefont{Furdyna}},
  \bibinfo{author}{\bibfnamefont{M.}~\bibnamefont{Grimsditch}},
  \bibnamefont{and} \bibinfo{author}{\bibfnamefont{R.}~\bibnamefont{Merlin}},
  \bibinfo{journal}{Phys. Rev. B} \textbf{\bibinfo{volume}{75}},
  \bibinfo{pages}{233308} (\bibinfo{year}{2007}).

\bibitem[{\citenamefont{Werpachowska and Dietl}(2010)}]{PRB82_85204}
\bibinfo{author}{\bibfnamefont{A.}~\bibnamefont{Werpachowska}}
  \bibnamefont{and} \bibinfo{author}{\bibfnamefont{T.}~\bibnamefont{Dietl}},
  \bibinfo{journal}{Phys. Rev. B} \textbf{\bibinfo{volume}{82}},
  \bibinfo{pages}{085204} (\bibinfo{year}{2010}).

\bibitem[{\citenamefont{Qi et~al.}(2007)\citenamefont{Qi, Xu, Tolk, Liu,
  Furdyna, and Perakis}}]{APL91_3}
\bibinfo{author}{\bibfnamefont{J.}~\bibnamefont{Qi}},
  \bibinfo{author}{\bibfnamefont{Y.}~\bibnamefont{Xu}},
  \bibinfo{author}{\bibfnamefont{N.~H.} \bibnamefont{Tolk}},
  \bibinfo{author}{\bibfnamefont{X.}~\bibnamefont{Liu}},
  \bibinfo{author}{\bibfnamefont{J.~K.} \bibnamefont{Furdyna}},
  \bibnamefont{and} \bibinfo{author}{\bibfnamefont{I.~E.}
  \bibnamefont{Perakis}}, \bibinfo{journal}{Appl. Phys. Lett.}
  \textbf{\bibinfo{volume}{91}}, \bibinfo{pages}{112506}
  (\bibinfo{year}{2007}).

\bibitem[{\citenamefont{Wirthmann et~al.}(2008)\citenamefont{Wirthmann, Hui,
  Mecking, Gui, Chakraborty, Hu, Reinwald, Sch{\"u}ller, and
  Wegscheider}}]{APL92_3}
\bibinfo{author}{\bibfnamefont{A.}~\bibnamefont{Wirthmann}},
  \bibinfo{author}{\bibfnamefont{X.}~\bibnamefont{Hui}},
  \bibinfo{author}{\bibfnamefont{N.}~\bibnamefont{Mecking}},
  \bibinfo{author}{\bibfnamefont{Y.~S.} \bibnamefont{Gui}},
  \bibinfo{author}{\bibfnamefont{T.}~\bibnamefont{Chakraborty}},
  \bibinfo{author}{\bibfnamefont{C.-M.} \bibnamefont{Hu}},
  \bibinfo{author}{\bibfnamefont{M.}~\bibnamefont{Reinwald}},
  \bibinfo{author}{\bibfnamefont{C.}~\bibnamefont{Sch{\"u}ller}},
  \bibnamefont{and}
  \bibinfo{author}{\bibfnamefont{W.}~\bibnamefont{Wegscheider}},
  \bibinfo{journal}{Appl. Phys. Lett.} \textbf{\bibinfo{volume}{92}},
  \bibinfo{pages}{232106} (\bibinfo{year}{2008}).

\bibitem[{\citenamefont{Khazen et~al.}(2008{\natexlab{b}})\citenamefont{Khazen,
  von Bardeleben, Cubukcu, Cantin, Novak, Olejnik, Cukr, Thevenard, and
  Lema{\^i}tre}}]{PRB78_195210}
\bibinfo{author}{\bibfnamefont{K.}~\bibnamefont{Khazen}},
  \bibinfo{author}{\bibfnamefont{H.~J.} \bibnamefont{von Bardeleben}},
  \bibinfo{author}{\bibfnamefont{M.}~\bibnamefont{Cubukcu}},
  \bibinfo{author}{\bibfnamefont{J.~L.} \bibnamefont{Cantin}},
  \bibinfo{author}{\bibfnamefont{V.}~\bibnamefont{Novak}},
  \bibinfo{author}{\bibfnamefont{K.}~\bibnamefont{Olejnik}},
  \bibinfo{author}{\bibfnamefont{M.}~\bibnamefont{Cukr}},
  \bibinfo{author}{\bibfnamefont{L.}~\bibnamefont{Thevenard}},
  \bibnamefont{and}
  \bibinfo{author}{\bibfnamefont{A.}~\bibnamefont{Lema{\^i}tre}},
  \bibinfo{journal}{Phys. Rev. B} \textbf{\bibinfo{volume}{78}},
  \bibinfo{pages}{195210} (\bibinfo{year}{2008}{\natexlab{b}}).

\bibitem[{\citenamefont{Sinova et~al.}(2004)\citenamefont{Sinova, Jungwirth,
  Liu, Sasaki, Furdyna, Atkinson, and {MacDonald}}}]{PRB69_85209}
\bibinfo{author}{\bibfnamefont{J.}~\bibnamefont{Sinova}},
  \bibinfo{author}{\bibfnamefont{T.}~\bibnamefont{Jungwirth}},
  \bibinfo{author}{\bibfnamefont{X.}~\bibnamefont{Liu}},
  \bibinfo{author}{\bibfnamefont{Y.}~\bibnamefont{Sasaki}},
  \bibinfo{author}{\bibfnamefont{J.~K.} \bibnamefont{Furdyna}},
  \bibinfo{author}{\bibfnamefont{W.~A.} \bibnamefont{Atkinson}},
  \bibnamefont{and} \bibinfo{author}{\bibfnamefont{A.~H.}
  \bibnamefont{{MacDonald}}}, \bibinfo{journal}{Phys. Rev. B}
  \textbf{\bibinfo{volume}{69}}, \bibinfo{pages}{085209}
  (\bibinfo{year}{2004}).

\end{thebibliography}

\end{document}